\documentclass[nopreprintline,12pt]{elsarticle}

\usepackage{amssymb}

\usepackage{url}
\usepackage{amstext}
\usepackage{amsmath}
\allowdisplaybreaks
\usepackage{tabularx}
\usepackage{booktabs}
\usepackage{array}
\usepackage{xcolor}
\usepackage{algorithm}
\usepackage{algpseudocode}

\journal{Computer Methods in Applied Mechanics and Engineering}

\begin{document}

\begin{frontmatter}



\title{Optimization of the Shape of a Hydrokinetic Turbine's Draft Tube and Hub Assembly Using Design-by-Morphing with Bayesian Optimization}


\author[inst1]{Haris Moazam Sheikh \corref{cor1}}
\author[inst1]{Tess A. Callan}
\author[inst1]{Kealan J. Hennessy}
\author[inst1]{Philip S. Marcus}

\affiliation[inst1]{organization={Mechanical Engineering Department, University of California, Berkeley},
            addressline={6116 Etcheverry Hall}, 
            city={Berkeley},
            postcode={94720-1740}, 
            state={California},
            country={U.S.A.}}

\cortext[cor1]{Corresponding author contact: \texttt{harissheikh@berkeley.edu}}

\begin{abstract}
Finding the optimal design of a hydrodynamic or aerodynamic surface is often impossible due to the expense of evaluating the cost functions (say, with computational fluid dynamics) needed to determine the performances of the flows that the surface controls. In addition, inherent limitations of the design space itself due to imposed geometric constraints, conventional parameterization methods, and user bias can 
restrict {\it all} of the designs within a chosen design space regardless of whether  traditional optimization methods or newer, data-driven design algorithms with machine learning are used to search the design space.  
We present a 2-pronged attack to address these difficulties: we propose (1) a methodology to create the design space using morphing that we  call {\it Design-by-Morphing} (DbM); and (2) an optimization algorithm to search that space that uses a novel Bayesian Optimization (BO) strategy that we call {\it Mixed variable, Multi-Objective Bayesian Optimization} (MixMOBO). We apply this shape optimization strategy to maximize the power output of a hydrokinetic turbine. Applying these two strategies in tandem, we demonstrate that we can create a novel, geometrically-unconstrained, design space of a draft tube and hub shape and then optimize them simultaneously with a {\it minimum} number of cost function calls. Our framework is versatile  and can be applied to the shape optimization of a variety of fluid problems.
\end{abstract}




\end{frontmatter}


\section{Introduction}
\label{sec:intro}


\subsection{Motivation and Background}
Renewable energy is fundamental to meeting our energy demands in a sustainable fashion. Although there exist several renewable sources for clean energy (wind, wave, solar, etc.), hydroelectric power is perhaps the most pertinent as a viable replacement for petroleum, natural gas and other fossil fuels. In fact, hydropower was one of the largest renewable energy sources in the United States in 2021 \cite{eiaenergyreview2021}. As such, it is important to consider ways in which the hydroelectric power plant (HEPP) can be more efficient, and thereby more cost-effective.
\par
The relative size of HEPP’s is usually classified by the amount of power they generate. Although there exists no unique definition of “small” hydropower, it is often accepted as a generating capacity at or below 10 megawatt electrical (MWe) \cite{smallhydro_models}. The demand for ``small'' hydropower is steadily increasing \cite{smallhydro_demand} despite concerns about its potential adverse environmental impacts (but impact studies and  how the impacts scale with the size of the hydropower plant remain controversial \cite{BAKKEN2012185, smallhydro_ecology}). Currently, there are numerous  undeveloped sites across the globe that have large potentials for efficient and sustained power generation via small hydropower \cite{smallhydro_idaho}. The sites with the most potential and that are easiest to exploit are those with low-impact stream-reaches, existing non-powered dams, and sites with existing conduits \cite{smallhydro_models}. 
Motivated by this potential for inexpensive and sustainable energy,   we propose here a new methodology for the design of some of the parts of a small hydrokinetic turbine. As a specific demonstration of our methodology, we provide a new, more efficient design of the draft tube and hub assembly of a small, low-head Kaplan turbine.
\par
The turbine component of the HEPP works by converting the potential and kinetic energy of the water entering the turbine into mechanical work, and then producing electricity via a generator \cite{reneng}. As one of the oldest and largest sources of renewable energy, there exists a wide variety of hydrokinetic turbines, but a concept common to all of them is the {\it dynamic pressure} or  {\it hydraulic head}  $P + \rho {\bf v}^2/2$ (where $P$ is the static pressure, $\rho$ the density and ${\bf v}$ the velocity) of the water entering the turbine. Using Bernoulli's principle (see $\S$~2) we can relate this head to the amount of power an ideal turbine can produce and the physical properties of the hydropower system such as the relative heights of the dam, penstock, and tail water discharge \cite{bernoulli}. For high head ranges, impulse turbines (e.g., Pelton)  exploit only the velocity of the fluid across the runner (see Fig. \ref{turbineschematic}) to create the mechanical of the turbine blade; whereas in medium and low head ranges, reaction turbines (e.g., Francis and, with later developments for low head applications, Kaplan) exploit both the fluid's velocity {\it and} the fluid pressure or enthalpy across the runner. It is the conversion of enthalpy that allows the draft tube and hub assembly to enhance the performance of a reaction turbine \cite{sheldon, sheldonopt}. The draft tube and hub assembly make up only part of the overall hydropower system's performance, but they are most important for low-head turbines, which we consider here to be those under 20~m.


\par
The {draft tube} is a diffuser, or several diffusers joined together, that sits beneath the runner of the turbine and directs the flow downstream of the turbine blades to the tailwater pool. It therefore has a large role in determining the dimensions of the lower section of the power plant \cite{cervantesthesis}. The draft tube increase the efficiency of the turbine by adjusting the dynamic pressure $(P + \rho {\bf v}^2/2)$ such that the static pressure $P$ just downstream of  the turbine blade is decreased. The adjustment is done by decelerating the velocity of the fluid passing through the draft tube, and it is this property (hereafter referred to as the ``pressure recovery'') that affects the power-generating capacity of Kaplan and other reaction turbines \cite{krivchenko, anderssonthesis}.
\par
The {\it hub} is a conically shaped part that extends just past the inlet of the draft tube, centered about the axis turbine's of rotation and connecting its blades. Water flows into the draft tube through an annular-shaped region between the hub and the draft tube wall. The hub rotates with the same angular velocity as the turbine blade, while the outer boundary of the draft tube is non-rotating. (See Fig.~\ref{turbineschematic}.) It is important to optimize the hub because it modifies the inlet flow to the draft tube, and therefore has a large  effect on the pressure recovery.
\par
Poorly designed parts of the draft tube or hub can significantly decrease a turbine's efficiency by exacerbating turbulence and increasing friction losses. 
For example, the draft tube's elbow (Fig.\ref{turbineschematic}), which is necessary for redirecting the tailwater flow, can promote flow separation due to excessive centrifugal force at its inner radius. Similarly, a poorly designed  hub can allow the swirl flow at the inlet due to the turbine blades  create instabilities in the flow that lead to noise, vibration (prompting failure due to fatigue), and even the reversal of flow through the center of the draft tube (causing sudden changes in power output of the turbine) \cite{nilsson1}. A well-designed hub can help prevent these problems, and, in addition,  allow a larger opening angle of the diffuser (and therefore higher pressure recovery). As such, we not only optimize the draft tube, but also the hub, which is usually neglected in prior optimization studies  \cite{marjavaarathesis, dahlhaug, gubin}.
\par
Much of the optimization effort of Kaplan draft tubes in recent years was focused on improving the sharp-heeled elbow draft tube \cite{gubin}. This shape was first installed in 1949 at the Hölleforsen Hydro Power Station in Vattenfall, northern Sweden, utilising a 25m head and with a power generating capacity of 150 MWe \cite{holleforsen_2022}. Gubin \cite{gubin} and Dahlbäck \cite{dahlback} independently argued that this draft tube shape needed improvement, and subsequently there have been many proposed design changes. \citet{marjavaaralundstrom} and \citet{marjavaarathesis} use a Response Surface Method (RSM) surrogate modeling strategy, as well as a commercial CFD code (ANSYS CFX4.4) to create new designs with different parametrizations (circular and elliptical, respectively) of the elbow section. More recent improvements include those of 
\citet{daniels_1, daniels_2} who use multi-objective Bayesian optimization to maximize pressure recovery using a series of subdividing curves, optimizing over the inflow cone, outer-heel, and secondary straight diffuser. Other improved designs focus on the optimization of the draft tube for low-head applications while retaining the sharp-heel shape \cite{eisinger, mcnabb}.
\par
%

\par

\subsection{Design-by-Morphing}
Design search space creation or parametrization is an integral part of any shape optimization process. Parameterization determines both the design space and the complexity of the optimization problem. A desirable parameterization technique must cover a wide design space within a limited number of design parameters\cite{sobester_barrett_2008,sripawadkul_comparison_2010,masters_review_2015}, necessary during the early design stage when minimum geometric constraints are placed and radical changes during the optimization process are required.

For a set number of parameters, parametrization techniques are judged on the different fidelity and ranges of control they offer \cite{moazam2017computational,masters_review_2015,sobester_barrett_2008}. Discrete methods \cite{discrete_method}, where design parameters are exactly the discrete surface points that define the airfoil shapes. The displacement of each point can be adjusted \cite{survey_parameterization}, and very fine local control can be achieved. However, to describe a shape accurately and with high fidelity, a large number of surface points are needed, which increases the number of parameters of the optimization problem. To accommodate the large number of design variables, gradient-based method are usually used to guide the optimization which can easily get stuck at a local optimum.

High-fidelity features can be captured by B-splines \cite{sanaye_multi-objective_2014, han_adaptive_2014}, and nonuniform rational B-spline (NURBS) \cite{nurbs} which form curves by connecting low-order B{\'e}zier segments defined by the control points. The fidelity of shape representation with these techniques depend on the number of parameters used for curve definition, increasing the computational complexity. To reduce the number of the design parameters, the control points can be grouped together, free-form deformation (FFD) method \cite{FFD1,FFD2}. Other techniques include spectral representation techniques which use some basis functions or modes, such as proper orthogonal decomposition (POD) \cite{POD1,POD2}, Hicks-Henne's approach \cite{hicks_wing_1978}, and class/shape function transformation (CST) method \cite{kulfan_fundamental_2006, akram_cfd_2021}.

To address the design challenges for improved hydrodynamic and aerodynamic surfaces, we introduced a new framework, Design-by-Morphing (DbM),  for creating a design space that is versatile enough to include old and new designs and that is sufficiently free of human bias to produce a radical and counter-intuitive design search space. The strategy cannot be called fully free from designer bias due to the choice of baseline shapes by designer, but the negative weights and radical baseline shapes, described later, minimize this effect. DbM uses the shape space as the basis for the parametrization scheme, thus allowing infinite fidelity without increasing the design space parameters for any shape, since it is the weight on the shape itself that is optimized. This is not possible with any conventional scheme discussed above and DbM has been shown to reproduce the entire UIUC database to very fidelity using just 25 parameters in \citep{airfoils}. DbM was first used by Oh et al. \cite{2018CompM6223O} and later used in other optimization problems \cite{airfoils,2019APS..DFDQ14007S}. An $N$-dimensional design space for a shape is created by choosing $N$ baseline shapes. The shape within our design space is specified by the choice of the $N$ weights of these baselines from which the design is morphed. The bounds of the  weights are sufficiently large that the morphed shapes can be not only interpolations among the shapes, but also extrapolations. Furthermore, any of the weights can be negative so that features of poorly performing baseline shapes can either be suppressed or entirely avoided. Negative weights and large positive weights allow for unintuitive shapes and for extrapolations.  Many design techniques only allow for small departures from existing designs \cite{computation6010005} or allow only local changes at one or a few specific locations, rather than global changes to the overall shape \cite{2007IJNME..71..313C, 1995IJNME..38.2283Z}.  This is especially problematic with most methods that use control points. CAD, and/or NURBS. 
Generally, methods with parametric control \cite{moazam2017computational,2007IJNME..71..313C, 2005CMAME.194.4135H, 2001PrAeS..37...59S,2011CMAME.200..883W} and the adjoint method \cite{computation6010005, 1995IJNME..38.2283Z, FANG1329} limit the amount of change that can be made to a design so that radical new designs are not possible. DbM  also allows a more extensive design space where both spatially local and global changes can be made to the  design, and those changes can be subtle and/or radical \cite{airfoils}. This allows us to find an optimum which may be a completely new, unconventional shape \cite{2018CompM6223O}. 

\subsection{Improved Bayesian Optimization}
Once a design space is chosen, many  engineering optimization problems require the repeated numerical (or laboratory) evaluation of an expensive multi-modal black-box  function to determine the performance or cost of a particular design. In optimizations of a surface that interacts with a fluid, generally the fluid flow must be computed with an expensive Computational Fluid Dynamics (CFD) program to determine the quantitative performance of each candidate design. Those quantitative results are then fed into an optimizing algorithm to determine the best performing design. The expense of computing the performance function with CFD makes many optimization problems intractable. Furthermore, the highly nonlinear behaviour of fluid flows often leads to a design space where the performance function has many {\it local}  maxima, and it is difficult for the search algorithm to find the {\it global} maximum. 

Optimizing draft tubes is an example of a search requiring an expensive multi-modal, black-box, performance function because the efficiency and pressure recovery of each candidate draft tube must be computed with CFD. Gradient based optimization techniques are unsuitable for multi-modal optimization since they get stuck in local optima and might require a number of function calls to either multi-start or get out of the local optimum. Global optimization schemes, such as evolutionary algorithm on the other hand require a large number of function calls to reach a global optimum, making them unsuitable for expensive black-box functions. Compared to Bayesian optimization, a genetic algorithm requires at least two orders of magnitude greater number of function calls to find the global optimum \cite{mixmobo}. Bayesian Optimization (BO) is an efficient search method for this type of optimization because it requires far fewer evaluations of the performance function to find an optimum than most other optimization algorithms \cite{brochu2010tutorial,williams2006gaussian}. BO techniques have been successful in the design of architected meta-materials \cite{frazier2015, chen2018computational, chen2019stiff, shaw2019computationally, song2019topology,sadvours}, hyper-parameter tuning for machine learning algorithms \cite{snoek2012practical, alphago2018, pmlr-v80-oh18a}, drug design \cite{articledrug, pmlr-v108-korovina20a}, and controller sensor placement \cite{JMLR:v9:krause08a}. In this study, we use an improvement that we made to BO, that we call  Mixed-variable Multi-Objective Bayesian Optimization (MixMOBO) \cite{mixmobo}. Previously, we used MixMOBO to design an architected meta-material that has  maximum strain-energy density \cite{sadvours}.

\section{Preliminaries}
\label{sec:prelim}

\begin{center}
\begin{figure}[!htb]
\centering
\includegraphics[width=0.9\linewidth]{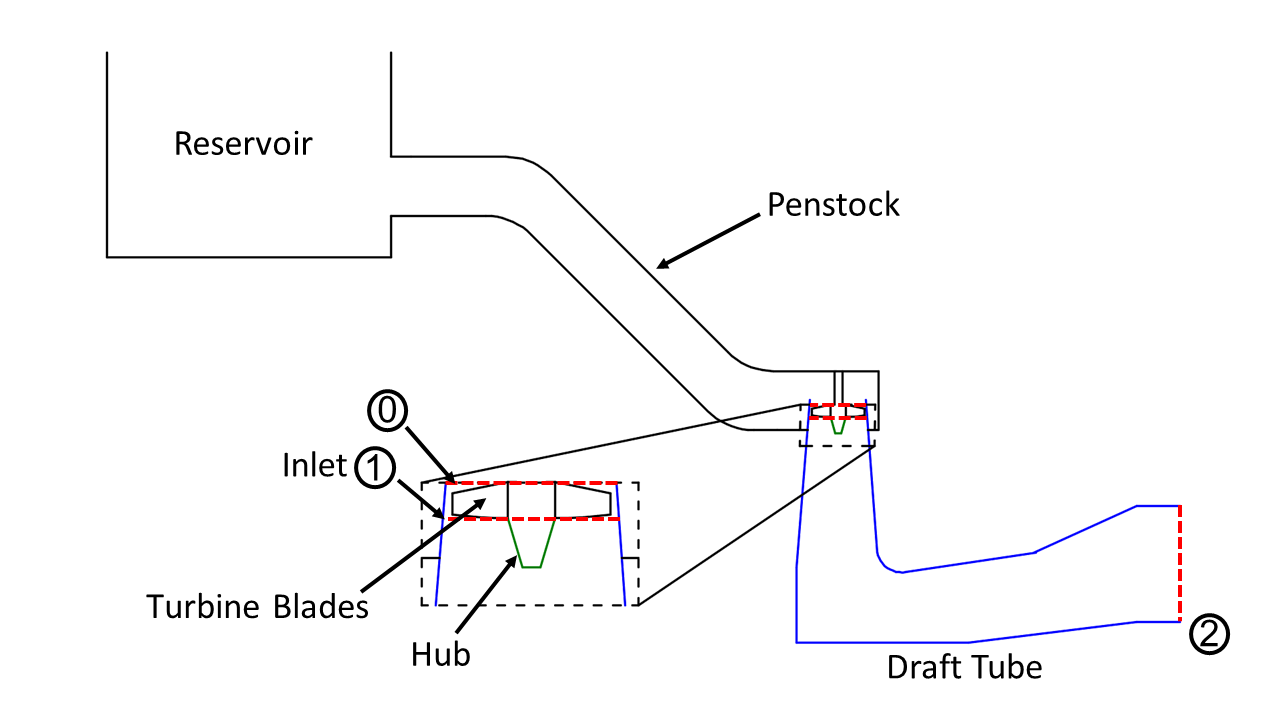} 
 \caption{Simplified schematic of a hydroelectric power plant. The enumerated cross-sections in red are: ({\bf 0}) the entrance to the turbine;  ({\bf 1}) the inlet of the hub/draft tube assembly; and ({\bf 2}) the outlet of the draft tube. The relative size of the draft tube (depicted in red) and hub (depicted in green) has been enlarged for clarity. The turbine blades are  between points ({\bf 0}) and ({\bf 1}). The reservoir surface is open to atmosphere. The pressure at the draft tube exit ({\bf 2}) is $P_2$ and is a given, fixed reference or gauge pressure.}
 \label{turbineschematic}
\end{figure}
\end{center}

Any kinetic energy or potential energy that is not converted into mechanical energy of the turbine shaft (and then into electrical energy) 
is discarded (i.e., wasted) when it leaves the turbine blades. A well-designed draft tube minimizes the wasted energy by
%
%
converting dynamic head into static head. The schematic of a hydrokinetic turbine assembly is shown in Figure~\ref{turbineschematic}. The overall drop in pressure across the turbine blade is  $(\Delta P) \equiv P_{0} - P_1$, where $P_{0}$ is the average fluid pressure just upstream of the turbine blade and is assumed to be fixed and independent of the designs of the hub and draft tube. $P_1$ is the average fluid pressure just downstream of the turbine blade, which is dependent on the designs of the hub and draft, and, in general, must be computed numerically and cannot be estimated using control volume analysis or a Bernoulli equation. However, control volume analysis does allow us to estimate  the theoretically available power, $\dot{W}$, that can  drive the turbine blade and produce electricity. It is
\begin{equation}
    \dot{W}=A_0 \, v_{0} \,  (\Delta P)
    \label{w4}
\end{equation}
where $A_0=A_1$ are the cross-sectional areas of the flow upstream and downstream of the turbine blade (where the subscripts refer to the location in Fig.~\ref{turbineschematic}) and $v_{0} = v_1$ are the characteristic velocities at these same locations.\footnote{Although $v_{0}$  can be thought of as an average streamwise velocity of the fluid upstream of the turbine blades, the amount of power that the blades can extract depends on the detailed flow interaction of the fluid velocity with the blade, which among other things depends on the swirl of the upstream velocity. Therefore, we leave the definition of this `characteristic' velocity purposefully undefined.} We assume that along with $P_0$, $A_0 = A_1$ and $v_0 = v_1$ are fixed, given parameters, and are not affected by the hub and draft tube designs. The goal of a well-designed hub assembly and draft tube is to maximize $\dot{W}$. 




Traditionally, \cite{ cervantesthesis, anderssonthesis,turbine99_iii}, the {\it mean pressure recovery} is defined as ${P_2 - P_1}$ and is the performance function used to the power output $\dot{W}$. We define the {\it dimensionless mean pressure recovery coefficient} as:

\begin{equation}
C_{prm} \equiv \frac{P_2 - P_1}{\frac{1}{2} \rho v_1^{2}} \\
\label{w3}
\end{equation}
where $\rho$ is the density of the fluid,  $P_1$ and $P_2$ are the average pressures, averaged over their respective cross-sections, and the subscripts refer to the location in Fig.~\ref{turbineschematic}. Using this definition and eq.~(\ref{w4}), we see that  
$\dot{W}$ and $C_{prm}$ are related by 
\begin{equation}
    \dot{W}=A_0 \, v_{0} \,  \left[P_0-P_2+C_{prm}(\rho v_1^{2}/2)\right].
    \label{w5}
\end{equation}
Because $(A_0 \, v_{0} \,  \rho \, v_1^{2})$ is positive (and because $A_0$, $v_{0}$,  $\rho$, $P_0$, $P_2$, and $v_1$ are assumed to be fixed, and independent of the designs of the hub and draft tube), maximizing $C_{prm}$ and maximizing $\dot{W}$ are equivalent. We therefore have chosen $C_{prm}$ to be the performance function that is maximized in this study.

Note that in order to prevent back flow without a draft tube, $P_1$ would need  to be greater than or equal to the given reference or gauge pressure $P_2$. Because  $P_0$ is assumed to be given (and the same value with or without a draft tube), the maximum value of $\dot{W}$ {\it without a draft tube} would be constrained by
\begin{equation}
    \dot{W} \le A_0 \, v_{0} \,  \left(P_0-P_2\right).
    \label{w6}
\end{equation}
The draft tube allows $P_1 < P_2$, and therefore allows an extra amount of power, $A_0 \, v_{0} \, C_{prm}(\rho v_1^{2}/2)$, to be generated. Note that because $\rho$, $P_2$, and $v_1$ are assumed to be fixed, maximizing  $C_{prm}$ minimizes $P_1$. Also note that because the fluid must be discharged from the draft tube at (${\bf 2}$)  with a finite velocity, and therefore the discharged fluid must necessarily contain some kinetic energy that cannot be recovered or converted into shaft mechanical energy, $C_{prm}$ can never sufficiently maximized, and $P_1$ never sufficiently minimized (even in the inviscid case) to make the generating system  100\% efficient \cite{reneng,  bernoulli,sheldon}.

\section{Methodology}
\label{sec:method}

\begin{figure*}[htb!]
\centering
\makebox[0.8\textwidth]{\includegraphics[width=1.25\textwidth]{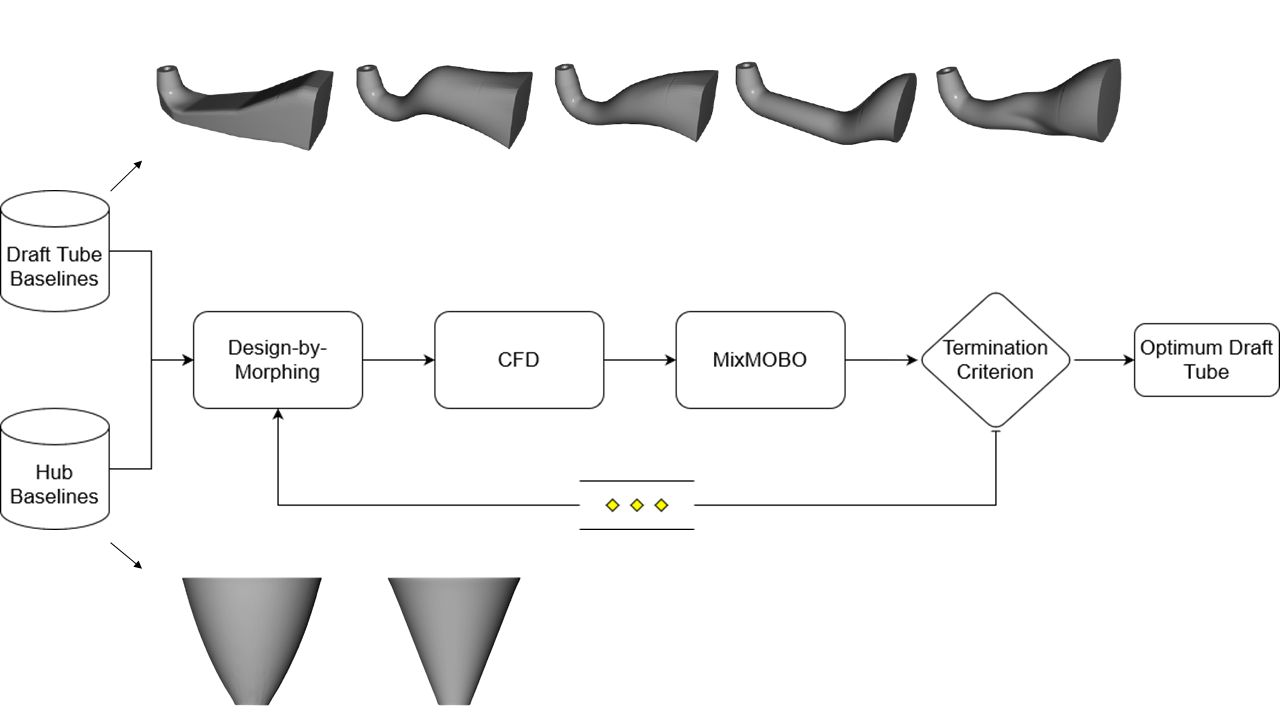}}
\caption{Optimization flowchart. The baselines for our draft tube and hub shapes are shown. The baselines are morphed together to create new shapes that are evaluated for their performance using CFD. This search space is sequentially optimized using parallel batches (represented by diamonds) of 5 shapes using MixMOBO, until the evaluation budget is reached. Note that the hub and the draft tube shapes are optimized simultaneously.} 
\label{fig:flow}
\end{figure*}

An overview of our procedure is demonstrated in the schematic in Fig.~\ref{fig:flow}. Our method starts with five different draft tube and two different hub baseline shapes to create the DbM search space. This search space is then initially sampled randomly for 50 data points. This data was then used to determine the next epoch or batch of designs or test points to evaluate using the MixMOBO algorithm. Each batch is a set of 5 data points or designs. As each new batch of designs is evaluated, their $C_{prm}$'s are added to the data base that MixMOBO uses to compute the next batch of designs to evaluate.This procedure continues until the evaluation budget is reached.
\par

\subsection{Baseline Shapes}
\label{sec:bshapes}

The design space is created from five baseline draft tube shapes and two baseline hub shapes. The shapes are  morphed with the weights chosen by MixMOBO to create a new draft tube and hub shape. These baselines are shown in Figure \ref{fig:flow}. The five baseline draft tube shapes are  homeomorphic to each other as are the two baseline hub shapes. Some of the baselines that we chose were used previously in literature so that we were able to exploit any inherent  advantageous features they might have, while other baselines were chosen to have non-intuitive features that would lead to radical designs, rather than incremental improvements. In the past, the fundamental design features of the Kaplan draft tube were formed through experimental observation and quasi-empirical formulae derived from geometries already installed and in use in HEPPs \cite{daniels_1}. The works of Gubin \cite{gubin}, Cervantes \cite{cervantesthesis},  Mulu \cite{muluthesis} and Nilsson et. al. \cite{nilsson1} \cite{nilsson2}) provide insights into draft tube geometries.
\par

Our first baseline shape for the draft tube is the sharp heel draft tube, which has been the focus of extensive optimization attempts. It is also the subject of extensive experimental and numerical studies, the majority of which were completed through the European Research Community On Flow, Turbulence And Combustion (ERCOFTAC) Turbine-99 Workshop series \cite{turbine99_i,turbine99_ii,turbine99_iii}. For this reason, this shape also provides excellent validation of our CFD setup. 
\par
The second and third baseline draft tube shapes are based on designs cited by \citet{gubin} and are particularly well-suited to  low-head applications (i.e., for Kaplan turbines). 
The fourth and fifth baseline tube shapes were both devised in order to create features which would expand the design space. These shapes were generated by taking the average shape of the first three baseline shapes and then adding radically different features such as a rounded outlet and a circular diffuser shape. The addition of these baseline shape can be thought of in a similar manner to adding mutation within a generation in genetic algorithms.
All the baseline draft tubes have the same shape at their inlets, and in all cases the planes containing inlet and outlet are perpendicular to each other. 

\par
Both baseline hubs have the same radius at the inlet, the same radius at the end of the hub, and the same length. The first baseline hub shape is based on one currently used in a Kaplan turbine \cite{muluthesis, MuluJonsson2012}. The second baseline hub is a cone shape, and which was historically used in a wide variety of low to medium head reaction turbines \cite{gubin}. Morphing these two hubs allows for a variation in the inlet geometry, which in turn modifies the flow profile near the inlet and therefore the pressure recovery. The baseline hub shapes are shown in Figure \ref{fig:flow}.
\par

\subsection{Design-By-Morphing}
\label{sec:dbm}

Design-by-Morphing (DbM) works by creating a one-to-one correspondence between the set of baseline shapes. The shapes are then ``morphed'' i.e. linearly combined together using ``weights''. These weights are the independent parameters that form the search space for optimizing the shape. DbM requires that the set of shapes be homeomorphic or topologically equivalent so that a one-to-one correspondence can be created between the surface elements. This one-to-one correspondence is created by defining a collocation strategy for the baseline shapes that ensures all constraints for our design are fulfilled (for example inlet and outlet orientation) even if individual shapes are radically different from each other.
\begin{figure}[!htbp]
\centering
\includegraphics[width=0.75\linewidth]{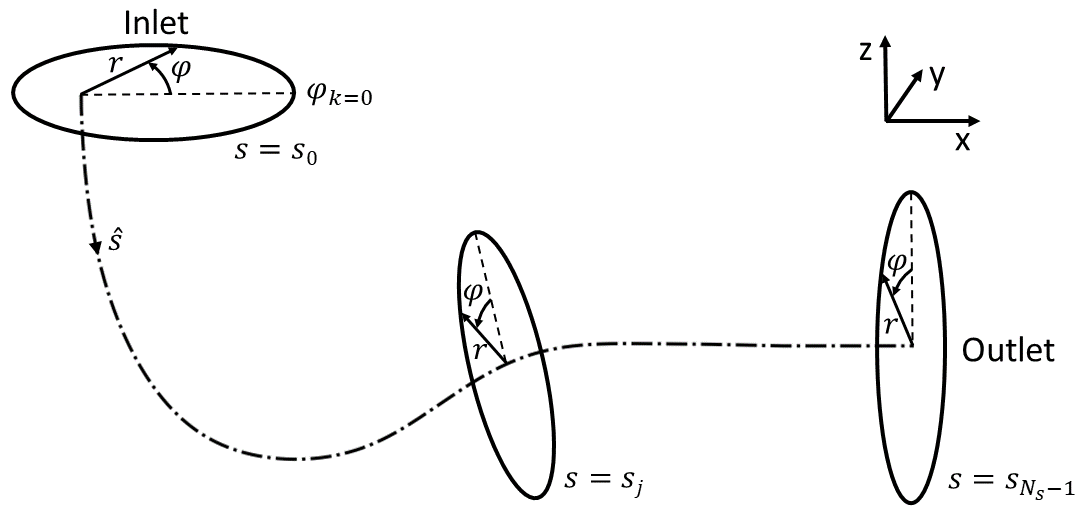}
 \caption{The {\it origin curve} is an arbitrarily chosen curve, shown in the figure with a dot-dash, that starts at the inlet and ends at the outlet. Its only constraint is that it must lie within all of the baseline draft tubes. Because all of the baseline draft tubes used in this study are reflection symmetric about the same $x$-$z$ plane, we chose to embed the origin curve within this symmetry plane. A unit vector $\hat{\bf s}$ lies along the origin curve pointing from the inlet to the outlet. The arc-length, or coordinate, along the origin curve is denoted as $s$, with a value of $s=0$ at the entrance and $s=S$ at the exit. We choose $N_s$ equally-spaced collocation points $\{s_i\}$, $i = 0, 1, 2, \cdots, N_s-1$ along the origin curve, with $s_0$ in the inlet plane, and $s_{N_s-1}$ in the outlet plane. Planes are defined at each value of $s_i$ that are locally perpendicular to the origin curve, with the planes at $s_0$ and $s_{N_s -1}$, at the inlet and outlet, respectively. The figure shows three of these planes as dashed lines at $s_0$, $s_j$, and $s_{N_j-1}$, where $1 \le j \le  N_s -2$.  The solid lines in each of these planes show the local polar coordinates$(r, \varphi)$ within each plane. In each of the $N_s$ planes, we chose the origin of $\varphi$ to lie in the $x$-$z$ symmetry plane, and the angle $\varphi =0$ is indicated in each of the three illustrated planes as broken lines.}
 \label{collocation}
\end{figure}

\begin{center}
\begin{figure*}[!htb]
\centering
\makebox[\textwidth][c]{\includegraphics[width=1.25\linewidth]{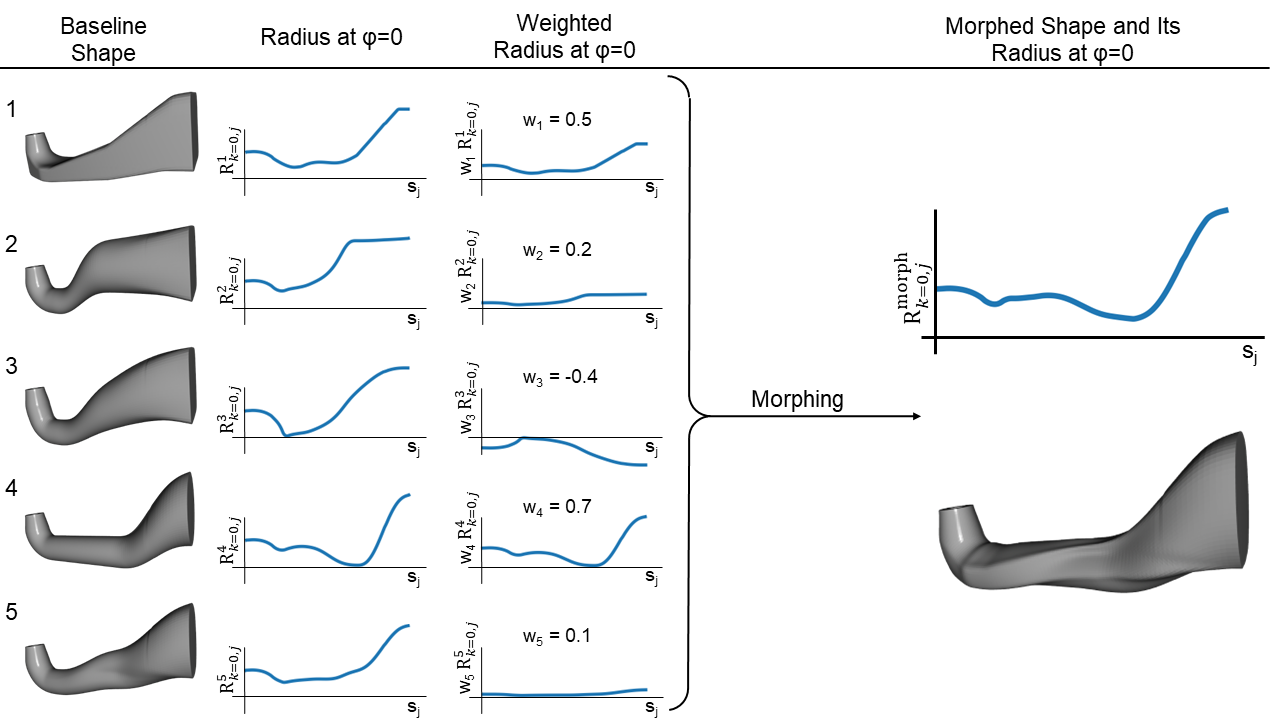}}
 \caption{An example of design-by-morphing (DbM). Column 1 shows the baseline shapes. Column 2 shows the radius $R^p_{k=0, j} \equiv r^p(\varphi_0 \equiv 0, s_j)$ of each baseline shape as a function of $j$. Note that for ease of visualization, we are only plotting the radius at the ``top'' of the baseline shape, rather than for all $\varphi$. Column 3 shows the weighted radius of each baseline at its top (i.e., the product of the baseline's weight with its radius at the top). Column 4 shows the morphed shape produced from the weights as well as the radius of the top of the morphed draft tube as a function of $s$.} 
 \label{dbm}
\end{figure*}
\end{center}

For the draft tube baseline shapes (shown in Figs.~\ref{collocation} and~\ref{dbm}), the inlet is constrained to a plane parallel to the horizontal ($x$-$y$) plane and the outlet is constrained to a plane parallel to the vertical ($y$-$z$) plane for all the baseline shapes. Note that this ensures that all the morphed shape inlets and outlets are similarly constrained i.e. inlets are in the same horizontal plane and the outlets are in the same vertical plane. Fig~\ref{collocation} explains how we construct a single coordinate system $(r, \varphi, s)$ for all of the baseline and morphed draft tubes that is mapped from the cylindrical coordinates $(r, \varphi, z)$. The  $z$ axis is mapped to the {\it origin curve} in the figure with with arc length $s$ and local unit vector $\hat{\bf s}$.  
The boundary of the draft tube is defined by its intersections with each of the $N_s$ perpendicular planes shown in Fig.~\ref{collocation}. Within each perpendicular plane, the intersection of the plane with the draft tube boundary is defined by the closed curve $r_j(\varphi)$, the radial distance of the draft tube boundary from the origin curve in the $j^{th}$ perpendicular plane. We note here that, because the same origin line is used for all of the baseline shapes, the start and end points of the origin line are always constrained to the same location in the inlet and outlet planes. The draft tube shape, however, changes around the origin line, meaning that the average path length for a fluid particle to travel from inlet to outlet, can vary for each draft tube.

In each of the $N_s$ perpendiclar planes, we discretize $\varphi$ into $N_{\varphi}$ equally-spaced collocation angles $\varphi_k$ with $k = 0, 1, 2,\cdots, N_{\varphi} -1$, with $\varphi_0 =0$ and $\varphi_{N_{\varphi}}  = 2 \pi (N_{\varphi} -1)/N_{\varphi}$.  
The radius of the $p^{th}$ draft tube as a function of the arc length $s$ and polar angle $\varphi$ is completely defined by the radial location matrix $R^p_{k, j} \equiv  r^p(\varphi_k, s_j)$, with $k = 0, 1, 2,\cdots, N_{\varphi} -1$, and $j = 0, 1, 2, \cdots, N_s -1$. Note that the $s_j$ perpendicular planes and the $\varphi_k$ collocation angles are the same for all of the the baseline and the morphed draft tubes.

For hub shapes, the same collocation strategy is used as we used for the draft tube shapes, with the origin curve of the hubs passing through the center of both of the baseline hubs because the baseline hub  are axi-symmetric. 

After morphing, each morphed radial matrix can be projected back into a 3-D shape as in Figure \ref{dbm}. 
\par
Because all of the five baseline draft tube shapes are  homeomorphic, they can easily be combined into a new morphed shape, given by $r^{morph}(\varphi_k, s_j) \equiv R^{morph}_{k, j}$, once the weights $w_{p}$, $p = 1, 2, \cdots, 5$ of each baseline are chosen (see Fig.~\ref{dbm}): 
\begin{equation}
    R^{morph}_{k, j} \equiv  \frac{1}{\sum_{p=1}^{5} w_{p}} \bigg|\sum_{p=1}^{5}\left[w_{p} R^p_{k, j}\right]\bigg|\quad \forall \; \sum_{p=1}^{5} w_{p} \neq 0. \label{shape}
\end{equation}
 
\par
Note that negatives weights and weights greater than unity allow for extrapolations; negative weights also allow us to ``avoid'' some baselines. This means that the only existing constraints are those imposed by choosing the baselines shapes themselves. That said, the lack of CAD parameterization means that DbM may produce non-physical, self-intersecting shapes when negative weights are applied. If the morphed shape is not physical because it has intersecting radial boundaries, we set the pressure recovery function of that morphed draft tube function to be zero so that the optimization method avoids that region of design space. For draft tube optimization, we limit the allowable range of the weights of each of the draft tubes such that: $w_p \in [-0.5,1.0]$.
\par
The morphed shape of the hub is given by an equation similar to eq.~(\ref{shape}), but the sum is over only 2 baseline radii. We denote the weights for hubs with $\alpha$ to differentiate it from the draft tube weights. Furthermore, the sum weight for the second hub baseline is constrained by $\sum_{l=1}^{2} \alpha_{l} =0.5.$, and for the morphed hub shape, $\alpha_1 \in [-0.5, 1]$. Due to the fact that sums of the weights of the draft tube baseline shapes and of the hub baseline shapes are fixed, there are only 6 degrees of freedom in choosing the values of the weights of the baseline shapes, so our design space has 6 dimensions.

\subsection{CFD Setup and Validation}
\label{sec:cfd}
\par
The coefficient of pressure recovery  $C_{prm}$ of each morphed draft tube shape that is physically allowable is determined with CFD. We validated the CFD code by comparing our simulations of the sharp heel draft tube with previously published results. In order to compute the flow in a morphed hub and draft tube shape, a mesh of that shape is created from a surface point cloud of the shape using Gmsh \cite{gmsh}. The average statistics of a typical mesh 
is given in Table \ref{meshStats}, where D is the inlet diameter. The software used is OpenFOAM, and the solver used is pimpleFoam.

\begin{table}[!ht]
\centering
\caption{Draft Tube Mesh Statistics}
\begin{tabular}[t]{p{0.5\linewidth}>{\raggedright\arraybackslash}p{0.3\linewidth}}
\toprule
Max Cell Size / D & 0.638 \\ 
Number of Elements & 7.9e+05 \\
Number of Nodes & 1.3e+05\\
Meshing Algorithm & Delaunay (3D) \\
\bottomrule
\end{tabular}
\label{meshStats}
\end{table}

\begin{figure}[!htbp]
\centering
\includegraphics[width=0.75\linewidth]{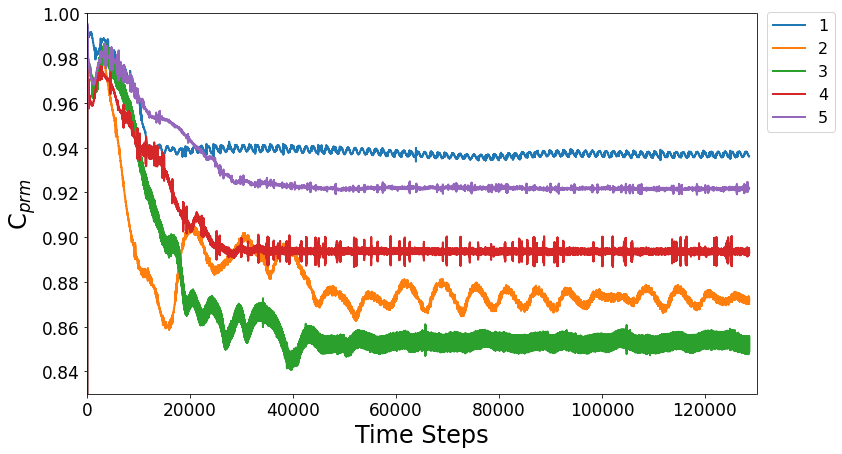}
 \caption{Numerical evaluation for all five baseline draft tube shapes using the first baseline hub shape. In our study, each simulation was run for 128,571 time steps. We chose this number of time steps to insure that the computed flow has come to a statistically steady state. Note that it requires approximately 50,000 time steps for the $C_{prm}$ to come to a statistically steady state.  (We {\it assume} that the flow itself comes to a statistically steady state some time between 50,000 and 128,571 time steps). The sharp heel draft tube (baseline shape {\bf 1}), has the highest $C_{prm}$ among the tube baseline shapes.}
 \label{time}
\end{figure}

For turbulence modeling, we use the  $k$-$\omega$~$SST$ turbulence model. This choice was made based on its success in previous draft tube studies \cite{anderssonthesis, wu}. Based on the kinematic viscosity of water, the  distance between the outer edge of the hub and the inlet wall, and the average streamwise velocity at the inlet, the  Reynolds number is 5.56e+05, and based on the azimuthal velocity of the hub, it is  9.48e+05. No-slip conditions are applied at the rotating, inner hub wall ($62.3~rad~s^{-1}$) and at the non-rotating draft tube boundary.\footnote{The optimization studies in the past for draft tubes \cite{eisinger, mcnabb}, did not consider a rotating hub inlet condition, which makes the CFD boundary conditions less close to the actual operating or experimental conditions} The inlet velocity is axisymmetric with non-zero values of the azimuthal, radial, and streamwise components. Our CFD simulations use the experimentally measured inlet velocities found by \citet{turbine99_ii}. The outlet pressure is the pressure at location ${\bf (2)}$ in Fig.~\ref{turbineschematic}, and is fixed gauge pressure, but the pressure at the inlet,  needed to determine the $C_{prm}$, is computed by the CFD code as a function of $r$, $\varphi$, and time. We run the CFD for 128,571 time steps for each morphed hub/draft tube shape. Our choice of the number of time steps is based on how long it takes the solution to reach a statistical steady state -- see Fig.~\ref{time}. 
Note that $C_{prm}$, given by eq.~(\ref{w3}), is not determined at a single time step, but rather it is averaged over the final 28,500 time steps of the computation (and note that as shown in Fig.~\ref{time}, the solution has converged to a statistically-steady equilibrium during those final time steps). The time step was chosen to be $0.00134~D/u_{avg}$, where $D$ is the outer diameter of the hub and $u_{avg}$ is the average streamwise inlet velocity. This time step was chosen based on a time resolution study (see below).

 The CFD is validated in two ways. The first compares our CFD-computed pressures with the experimentally-measured values along the top and bottom center lines of the baseline 1, Sharp-Heel, draft tube. The second uses baseline 1 to test the convergence of the CFD code by refining the numerical time step and grid size. We note here that the experimental studies did not provide the $C_{prm}$ values, thus no $C_{prm}$ comparison could be made.

\begin{figure}[!hb]
\centering
\includegraphics[width=0.75\linewidth]{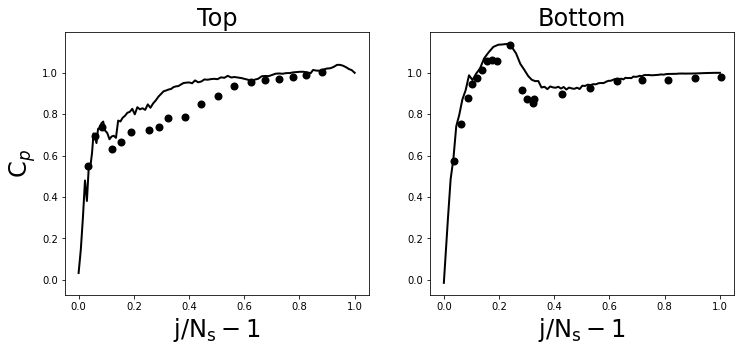}
 \caption{Pressure Recovery at the ``Top'' or ``Bottom'' of the Sharp-Heel draft tube and Hub 1 (in the $x$-$z$ symmetry plane). According to our collocation defined in Section~\ref{sec:dbm}, the ``Top'' corresponds to $R^1_{k=0,j}$ and ``Bottom'' corresponds to $R^1_{k=N_\varphi/2,j}$ as a function of j, where $j=0$ is at the inlet. We define $C_p(j)  \equiv \frac{P_{wall}(j)-P_{wall}(0)}{\frac{1}{2} \rho\left(\frac{Q}{A_1}\right)^{2}}$, where $P_{wall}(j)$ is the pressure at the draft tube wall. Solid circles are the experimental data; continuous curves are our numerically computations. The time step is $0.00134~D/u_{avg}$.}
 \label{TopAndBottomLine}
\end{figure}

\par
The local pressure recovery along the top and bottom lines of the first baseline shape (i.e., the sharp heel draft tube) is shown (and defined) in Figure \ref{TopAndBottomLine}. The figure compares our numerically-computed local pressure recovery values to the experimentally-measured values \cite{turbine99_ii}. Figures similar to Fig.~\ref{TopAndBottomLine} appear throughout the ERCOFTAC Turbine-99 Workshop series, and the deviation between our CFD and experimental values are consistent with the previous numerical studies. 

The second way we validated our use of our OpenFoam code and gridding scheme was by time step and mesh refinement. We chose our time step size by decreasing its value in the CFD code until the late time-averaged $C_{prm}$ values of the sharp heel draft tube with the first hub baseline shape changed by less than $0.01\%$. In particular, once we set the value of the time step as $0.00134~D/u_{avg}$, we repeated the calculation of the local pressure recovery $C_p(j)$ for both the top and bottom as a function of $j$ as defined in  Fig.~\ref{TopAndBottomLine} -- once with a time step $20\%$ greater than that used in the figure and once with it $20\%$ lower.  Plots of these new curves computed with these time steps are indistinguishable from the curves (error less than 0.5\%) shown in Fig.~\ref{TopAndBottomLine} (at the resolution at which we created the figure), validating that our time step is sufficiently small. We determined the size of the spatial resolution of the grid for the CFD in a similar manner: we decreased the grid size (or equivalently, increased the square of the number of grid points) until the late time-averaged values $C_{prm}$ values of the sharp heel draft tube changed by less than $0.4\%$. We also recomputed the two curves of the $C_p(j)$ in Fig.~\ref{TopAndBottomLine} with a spatial resolution of the spatial gridding mesh with a size that was $20\%$ greater than that used in the figure and once with it $20\%$ lower. Again, we found that the plots of these new curves computed with these grids are indistinguishable (error less than 0.5\%) from the curves shown in Fig.~\ref{TopAndBottomLine}, validating that our grid is sufficiently fine.

It is important to note that Fig.~\ref{time} shows that it takes more than 50,000 time steps for our numerically-computed $C_{prm}$ to reach a statistical equilibrium. The 50,000 time steps corresponds to a time that is approximately equal to 10 ``advective times'', where the latter is the streamwise length of the draft tube divided by the average streamwise velocity at the inlet $u_{avg}$. Previous numerical studies of draft tubes were often validated by comparing the numerically-computed local pressure recovery $C_{p}(j)$ at the bottom or top of the draft tube with the experimentally measured values. However, in many of those studies the codes were run for only {\it one} advective time (5000 of our time steps) before  the pressure recovery factor $C_{prm}$ was evaluated \cite{turbine99_ii}. Figure~\ref{time} clearly shows that although that time be sufficient to numerically capture the final statistically steady state of $C_{p}(j)$, it is insufficient for the full flow and  $C_{prm}$ to have come to a statistical steady state. 


\subsection{Bayesian Optimization}
\label{sec:bo}

In our previous work, we developed a Mixed-variable, Multi-Objective Bayesian Optimization (MixMOBO) algorithm \cite{mixmobo}, a framework for optimizing mixed-variable, multi-objective problems with noisy black-box function 
using parallel batch updates of the data set, that is, letting the acquisition function pick several points to evaluate at each iteration to allow their CFD evaluations to be carried out in parallel. MixMOBO was proven to be more efficient in terms of number of black-box function evaluations compared to other algorithms in small data settings. In our previous studies, we used MixMOBO to design a microlattice structure with the objective of maximizing its strain-energy density \cite{sadvours}. 
We are also applying MixMOBO for the optimization of vertical axis wind turbines \cite{2019APS..DFDQ14007S}. 

Our current study adapts the generalized MixMOBO algorithm for the specific case of 6 continuous variables with parallel batch updates using the HedgeMO algorithm \cite{mixmobo}. Our HedgeMO strategy uses the Upper Confidence Bound, Expected Improvement, Probability of Improvement, and Stochastic Monte-Carlo \cite{brochu2010tutorial, mixmobo} acquisition functions. We apply our Bayesian optimization algorithm in tandem with DbM here for the shape optimization of a draft tube/hub. This section details our adaptation of the MixMOBO algorithm for our current study.

The search  in our 6-dimensional design space (consisting of the 5 weights of the baseline draft tubes and the 1 independent weight of the two baseline hubs) for the draft tube/hub design with the maximum $C_{rpm}$ is an example of an optimization problem that can generically be posed as: 
\begin{equation}
\vec{w}_{opt}= argmax_{\vec{w} \in \mathcal{W}} [f(\vec{w})],
\end{equation}
where $f(\vec{w})$ is the objective to be maximized (in this case $C_{prm}$), and $\vec{w}$ is a $D$-dimensional (in this case 6-dimensional)  variable vector, defined over a bounded set $\mathcal{W}\subset \mathbb{R}^D$ (in this case, the weights of the baselines). For many practical engineering problems, $f(\vec{w})$ is expensive to evaluate (as is the case here, which requires repeated CFD evaluations of a complex draft tube designs). In such cases, a Bayesian optimization  algorithm is often the best choice \cite{brochu2010tutorial}. Bayesian optimization is a sequential optimization technique, specifically designed to find the optimal solution of a noisy black-box function $f$ {\it with the fewest possible evaluations or function calls} to $f$. At every iteration of the algorithm, a surrogate model $g$, usually a Gaussian process (GP), is fit over the data set $\mathcal{D}=\{[w_1,f(w_1)], \ldots,[w_i,f(w_i)]\}$.
\begin{equation}
{g}(\vec{w}) \sim GP\bigl({\mu ( \vec{w} ) },ker(\vec{w},\vec{w}')\bigr)
\end{equation}
Here $i$ is the total number of evaluated points (in this case, morphed draft tube/hub designs) and $\vec{w}$ is the vector of continuous variable vector. 

The MixMOBO algorithm is agnostic to the choice of kernel function. For our current study, we use a simple modified squared exponential kernel due to its good performance for continuous variable problems:

\begin{equation}
     ker({\vec{w}, \vec{w}'}) \equiv \epsilon _f ^2 \,\, exp \left [-\frac{1}{2}\lvert {\vec{w}}, {\vec{w}}'\rvert _{C}^T \,\, \,\, {\underline{\underline{ M}}} \,\, \,\, \lvert {\vec{w}},  \vec{w}'\rvert _{C}\right],
\label{eq:1}
\end{equation}

where $\vec{\theta} = (\{{\underline{\underline{M}}}\},\epsilon _f)$ is a vector containing all the hyper-parameters, $\{{\underline{\underline{M}}}\}=diag(\vec{h})^{-2}$ is the covariance hyper-parameter matrix and $\vec{h}$ is the vector of covariance lengths and distance between the variables defined to be their Euclidean distances.

Once the GP surface has been fit, the next query point in the design space that should be chosen to evaluation must explore the space that most probably contains the optimum (e.g., the design with the largest $C_{prm}$), and must also try to improve the fit for the next iteration, i.e. choose a region of space where information is sparse. These two determinations are not the same, and the former is called ``exploitation'' and the latter ``exploration''. Bayesian optimization strives to create a balance of exploitation and exploration that continues until there is either evidence that a global optimum has been found or a maximum prescribed number of iterations of the algorithm is reached. An acquisition function, which we represnt here by $\mathcal{A}^l$, determines the next query point (or points) $\vec{w}_{i+1}$ in the design space to be  evaluated with $f$ (in this case, with the costly CFD), by balancing the competing needs of exploitation and exploration. Once the next query point (or points) has been determined , $f$ is evaluated for that point (or points) and is then appended to the data set, $\mathcal{D}=\mathcal{D} \cup (w_{i+1},f(w_{i+1}))$. To optimize the exploitation/exploration balance, we use a GA to optimize the acquisition functions, which, although expensive to use on an actual black-box function, is an ideal candidate for optimizing the acquisition function working on the surrogate surface.

MixMOBO uses HedgeMO, a hedging strategy for acquisition functions, as part of its framework. Hedge strategies use a portfolio of acquisition functions, rather than a single acquisition function \cite{brochu2011portfolio}. HedgeMO also allows parallel `$Q$-batches' of query points. The detailed algorithmic flowchart of our adapted MixMOBO algorithm is shown in Algorithm \ref{alg:mixmobo}. We note that the convergence criterion presented by \citep{brochu2011portfolio} holds for our portfolio as well.

\begin{algorithm}[!ht]
  \caption{Continuous variable MixMOBO algorithm with HedgeMO}
  \label{alg:mixmobo}
\begin{algorithmic}[1]
  \State {\bfseries Input:} Black-box function $f(\vec{w}): {\vec{w} \in \mathcal{W}}$, initial data set size $N\_i$, batch points per epoch $Q$, total epochs $N$, mutation rate $\beta\in [0,1]$, parameter $\eta \in \mathbb{R}^{+}$
  \State {\bfseries Initialize:} Sample black-box function $f$ for  $\mathcal{D}=\left\{\left(\vec{w}_j,f(\vec{w}_j)\right)\right\}_{j=1:N\_i}$
\vspace{0.2cm}
  \For{$n=1$ {\bfseries to} $N$}
\vspace{0.15cm}  
  \State Fit a noisy GP surrogate function ${g}(\vec{w}) \sim GP\bigl ({\mu ( \vec{w} ) },ker(\vec{w},\vec{w}')\bigr)$ 
  \State For $L$ total acquisition functions, from each $\mathcal{A}^l$ acquisition function, propose $Q$-batch test-points, $\left\{(\vec{u})_n^l\right\}_{1:Q}=\left\{argmax_{\vec{u} \in \mathcal{W}}\mathcal{A}^l\left(g\right)\right\}_{1:Q}$ within the constrained search space $\mathcal{W}$ using multi-objective GA. A batch of $Q$ points is selected from this set using HedgeMO (steps 7-14)
  \State Mutate point $\left\{(\vec{u})_n^l\right\}_{q}$ within search space $\mathcal{W}$ with probability rate $\beta$ if $L_2$-norm of its difference with any other member in set $\left\{(\vec{u})_n^l\right\}_{1:Q}$ is below tolerance
  \vspace{0.15cm}  
\For{$l=1$ {\bfseries to} $L$}
  \vspace{0.15cm}
  \State For $l^{th}$ acquisition function, find rewards for $Q$-batch points nominated by that AF from epochs ${1{:}n}$-1, by sampling from $g$, $\left\{{\gamma}^l _ {1:n-1}\right\}_{1:Q} = \mu(\left\{(\vec{u})_{1:n-1}^{l}\right\}_{1:Q})$
  \vspace{0.15cm}
  \EndFor
  \vspace{0.2cm}
  \State Normalize rewards for each $l^{th}$ AF, $\phi_l=\sum_{j=1}^{n-1}\sum_{q=1}^{Q}\frac{\left\{{\gamma}^l _ {j}\right\}_{q}-min(\Gamma)}{max(\Gamma)-min(\Gamma)}$, where $\Gamma$ is defined as $\Gamma=\left\{{\gamma}^{1:L} _ {1:n-1}\right\}_{1:Q}$
  \State Calculate probability for selecting nominees from $l^{th}$ acquisition function, $p^l=\frac{exp(\eta\phi_l)}{\sum_{i=1}^L exp(\eta\phi_i)}$
  \vspace{0.2cm}
  \For{$q=1$ {\bfseries to} $Q$}
  \vspace{0.15cm}
  \State Select $q^{th}$ nominee as ${\vec{w}_n}^q$ from $l^{th}$ AF with probability $p^l$
    \vspace{0.15cm}
  \EndFor
  \vspace{0.2cm}
  \State Evaluate the selected points from the black-box function, $\{f(\vec{w}_n)\}_{1:Q}$
  \State Update $\mathcal{D}=\mathcal{D} \cup \left\{\left(\vec{w}_{n},f(\vec{w}_{n})\right)\right\}_{1:Q}$
  \vspace{0.15cm} 
  \EndFor
  \vspace{0.2cm}
\State {\bfseries return} Optimal solution set $\left\{\left(\vec{w}_{opt},f(\vec{w}_{opt})\right)\right\}$
\end{algorithmic}
\end{algorithm}

The process is repeated until  the evaluation budget is reached. Our choice for the evaluation budget was based on several optimization experiments on test functions and design spaces, described below, that we believe to be representative of the draft tube/hub optimization. To  determine the number of  iterations of the Bayesian optimization algorithm, or {\it epochs} (where each epoch determines the $C_{prm}$ of 5  morphed draft tube/hub designs in parallel), required to likely find the optimal design within the design space, we carried out optimizations of  a suite of test functions with different properties and whose maximum  values could be determined analytically. We optimized the Spherical, Rastringin, Syblinski-Tang and Amalgamated functions, all of which are standard functions used to test optimization schemes \cite{HAL} with the exception of the Amalgmated function, which is novel and created by us to mimic some of the properties of the draft tube/hub design space. The test functions that we used to determine the number of epochs that are needed for MixMOBO to likely find the maximum $C_{prm}$ are defined in Table~\ref{table_tf}. Similar to the design space of the draft tube/hub, each test function was tested with 6 dimensions.
\begin{table*}[!hbt]  
\label{table_tf}
\caption{Benchmark test functions}
  \centering
  \hspace{0.5cm}
  \resizebox{\textwidth}{!}{%
  \begin{tabular}{ccc}
    \toprule
   Name & Objective Functions & Notes\\\midrule
 $\text{Spherical}$ & $
  
\begin{aligned}
f(\vec{w}) &= -w_i^2,\: \ w_i\in(-10,10)\\[1ex]
\end{aligned}
$
& $\begin{aligned} &\text{convex} \\\end{aligned}$\\   
    $\text{Rastringin}$ & $
  
\begin{aligned}
f(\vec{w}) &= -[10+w_i^2-10\operatorname{cos}(2\pi w_i)],\:\ w_i\in(-5.12,5.12)\\[1ex]
\end{aligned}
$
& $\begin{aligned} &\text{non-convex}\\\end{aligned}$\\ \\

    $\text{Syblinski-Tang}$   & $
  
\begin{aligned}
f(\vec{w}) &= -\frac{w_i^4-16w_i^2+5w_i}{2},\:\ w_i\in(-5,5)\\[1ex]
\end{aligned}
$
& $\begin{aligned} &\text{non-convex} \\\end{aligned}$ \\ \\

    $\text{Amalgamated}$ & $
  
\begin{aligned}
f(\vec{w}) &=\sum_{i=1}^{D}-g(w_i)  \\
g(w_i)&=
\left\{
\begin{array}{@{\:}l@{}l}
    - \operatorname{sin}(w_i),\:\text{if}\:   k=0, \ w_i\in(0,\pi)\\[1ex]
    \frac{w_i^4-16w_i^2+5w_i}{2},\: \text{if} \:   k=1, \ w_i\in(-5,5)\\[1ex]
    w_i^2,\: \text{if}\:  k=2, \ w_i\in(-10,10)\\[1ex]
    [10+w_i^2-10\operatorname{cos}(2\pi w_i)],\:\text{if}\:  k=3, \ w_i\in(-5.12,5.12)\\[1ex]
    [100(w_i-w_{i-1}^2)^2+(1-w_i)^2],\:\text{if}\:  k=4, \ w_i\in(-2,2)\\[1ex]
    -|\operatorname{cos}(w_i)|,\:\text{if}\:  k=5, \ w_i\in(-\pi/2,\pi/2)\\[1ex]
    w_i,\:\text{if}\: k=6, \ w_i\in(-30,30)
\end{array}
\right. \\
k &= \operatorname{mod}(i-1,7), ~ i=1,...,n \\
\end{aligned}
$
& $\begin{aligned} &\text{non-convex,} \\ &\text{non-uniform,} \\ &\text{anisotropic}\end{aligned}$ \\ \\
    
    \\
    \bottomrule
    \end{tabular}}
    \caption{Details about the test functions, other than the  {\it Amalgamated function}, are given in  \citet{HAL}. All of the test functions have known global maxima. We created the  {\it Amalgamated function}, a piece-wise function formed from commonly used analytical test functions with different features.  The Amalgamated function is non-convex and anisotropic (as is the design space of the draft tube/hub), unlike the other test functions listed here, which are isotropic. These other test functions are commonly used for testing optimization algorithms. Similar to the optimization of the draft tube/hub, each test function here has 6 dimensions.}
    \label{tab:tests}
\end{table*}

\begin{figure}[h]
\centering
\includegraphics[width=0.75\linewidth]{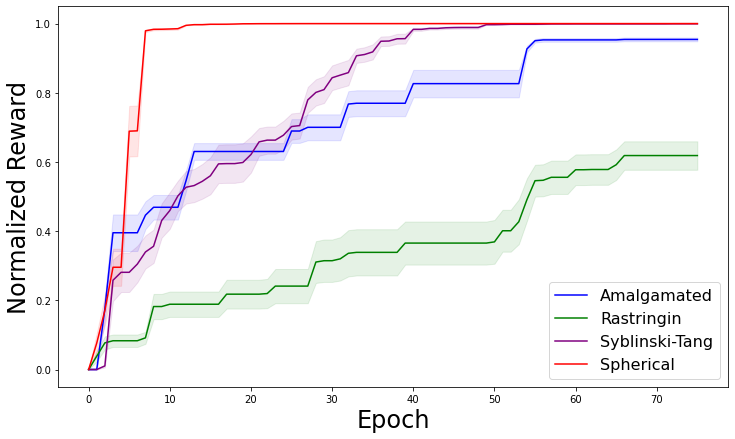}
 \caption{Benchmarks for determining the number of epochs of MixMOBO needed to find the likely global maximum. The number of random evaluations of the function to be maximize that are needed to initialize MixMOBO, was set at 50. Our MixMOBO algorithm was then run with 5 parallel batch evaluations per epoch. The evaluation budget was 425 black-box function evaluations including the 50 initial random evaluations. Thus, we ran the code for $75 = (425 - 50)/5$ epochs. Because each optimization begins with random evaluations, the algorithm was run 5 times for each test function to determine the average and standard deviation of the number epochs needed to find the global maximum. The mean Normalized Reward, defined as \textit{(current optimum-random sampling optimum)/(global optimum - random sampling optimum)} is plotted along with the 0.2*standard deviation plotted as colored bands. A Normalized Reward of unity means that the algorithm has successfully found the global maximum.}
 \label{benchfig}
\end{figure}

Figure~\ref{benchfig} shows how MixMOBO approaches the global maximum of each of the four test functions as a function of epoch. Two of the four test functions find the global maximum within 75 epochs.

The optimization was terminated after 75 epochs because the test functions reached reasonable convergence, other than Rastringin function, which is know to have a a multitude of local optima, as shown in Fig.~\ref{benchfig}. We could only afford 75 epochs worth of CFD evaluations of candidate draft tube designs in our computational budget so the evaluations were not carried past 75 epochs. As depicted in our Results, the draft tube optimization converged to the final design in 30 epochs. MixMOBO, like most Bayesian optimization schemes, needs to be initialized with evaluations of random designs. Based on the results shown on Test Functions, we use 50 random evaluations, and 75 epochs with 5 parallel batch evaluations per epoch to optimize our 6-dimensional design space.

\section{Results}
\label{sec:results}
\begin{figure*}[!htpb]
\centering
\includegraphics[width=1\linewidth]{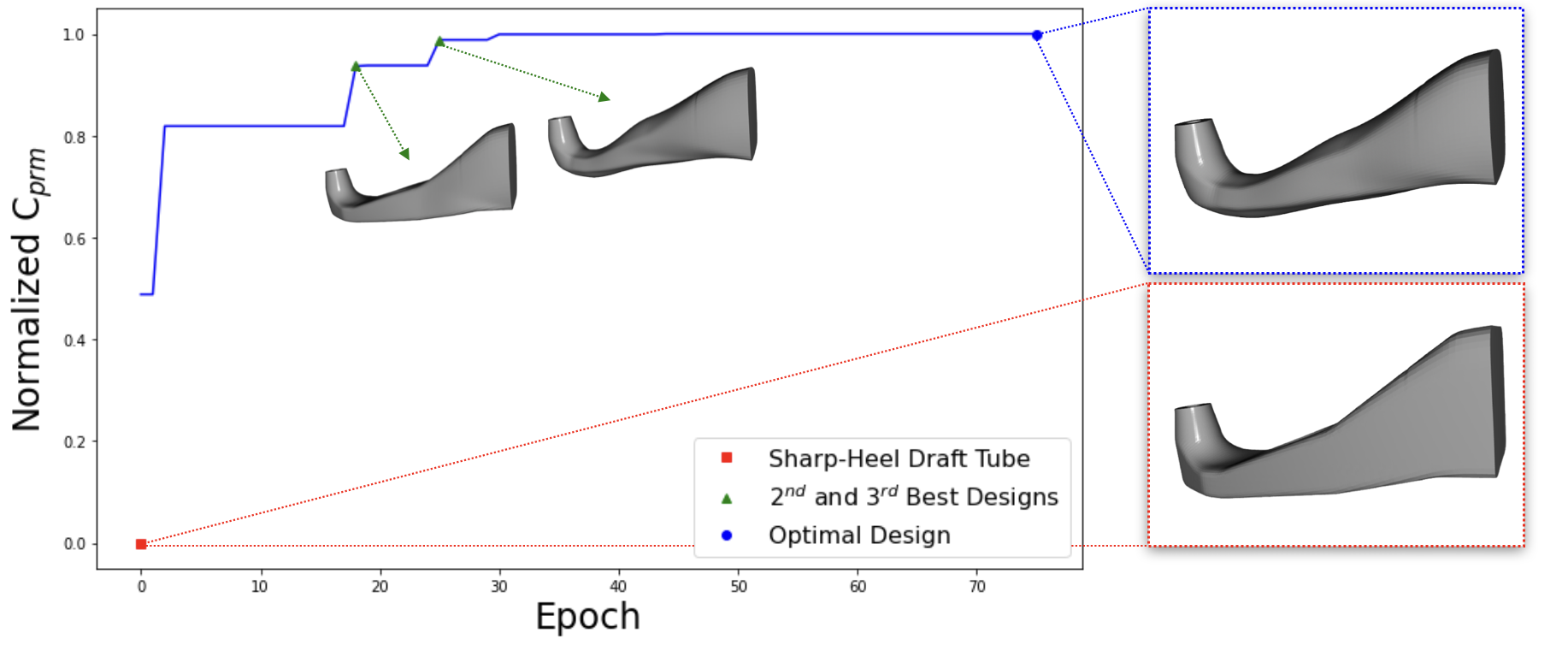}
 \caption{Optimization progress showing the Normalized $C_{prm}$ of the current epoch's optimum as a function of  the epoch number. The  Normalized $C_{prm}$ is defined in the text. By definition, the Normalized $C_{prm}$ must increase monotonically with the epoch number. The morphed draft tube/hub with the  second highest, and third highest $C_{prm}$'s are  shown  within the main figure. The designs that are illustrated in the panels to the right of the main figure are the shape of 
 the morphed draft tube/hub design with the highest $C_{prm}$ (framed with a blue box) and also the sharp heel draft tube with the first baseline hub  (framed with a red box). The main figure begins at epoch zero and the  Normalized $C_{prm}$ at epoch~0 is that of the design of 50 random designs used to initialize MixMOBO with the highest  $C_{prm}$.}
 \label{resultsfig}
\end{figure*}

We searched for the morphed draft tube/hub design with the maximum $C_{prm}$ in our 6-dimensional design space using MixMOBO with  50 initial random designs and 75 epochs  with 5 design evaluations per epoch. The results our shown in Fig.~\ref{resultsfig} and Table~\ref{weightstable}. The Normalized $C_{prm}$ used in the figure is defined as: $\Bigl[(C_{prm}$ of  Current Epoch's Optimum$) -  (C_{prm}$ of Sharp heel draft tube with the baseline~1 hub$)\Bigr] \Big/ \Bigl[(C_{prm}$ of the Optimum design found by MixMOBO$) - (C_{prm}$ of Sharp heel with baseline~1 hub$)\Bigr]$.

We note here that our CFD simulations are run to convergence, as explained in Section~\ref{sec:cfd}. If we ran our simulations for only one advective time, we would potentially have gotten a higher $C_{prm}$ value, which would not be converged and erroneous, as shown in Fig.~\ref{time}.

Figure~\ref{resultsfig} and the Table show that the  $C_{prm}$ of the sharp heel draft tube with the baseline hub~1 is significantly lower than the $C_{prm}$ of the best (and second and third best) design found with MixMOBO. In fact, the $C_{prm}$ of the best of the 50 {\it random} designs that were used to initialize MixMOBO was better than that of the sharp heel draft tube, which shows the strength of our Design-by-Morphing approach. The sharp heel draft tube had the best $C_{prm}$ (when coupled with hub shape 1) of all the baseline draft tubes. We note that the $C_{prm}$ of the optimal design found by MixMOBO is significantly better than the Sharp-Heel draft tube, which is the draft tube of choice for Kaplan turbines around the world \cite{gubin}.

\begin{table*}[htbp]
    \centering
        \begin{tabular}[t]{*{9}{c}}
        \toprule
        Design &Epoch& $\omega_1$ & $\omega_2$ & $\omega_3$ & $\omega_4$ & $\omega_5$ & $\alpha_1$ & $C_{prm}$ \\
        \midrule
        Sharp-Heel & & 1 & 0 & 0 & 0 & 0 & 0.5 & $\mathbf{0.9370}$ \\
        $3^{rd}$ Best  &19& 0.86 & -0.17 & 0.40 & -0.21 & 0.98 & 0.98 & $\mathbf{0.9607}$ \\
        $2^{rd}$ Best &26& 0.88 & -0.24 &0.26& -0.17 & 0.99 & 0.92 & $\mathbf{0.9617}$ \\
        Optimal &30& 0.90 & -0.14	& 0.22 & -0.25	& 0.99 & 0.99 & $\mathbf{0.9620}$ \\
        \bottomrule
        \end{tabular}
    \caption{$C_{prm}$ and DbM weights of draft tube/hub designs. The designs listed are those with the best, second best, and third best $C_{prm}$'s found with MixMOBO. We also show the epoch number when the design was found by MixMOBO. (See Fig.~\ref{resultsfig}.) In addition, the sharp heel draft tube with the baseline~1 hub is listed.}
    \label{weightstable}
\end{table*}



Note that all three of the morphed draft tube/hub designs in Table \ref{weightstable} have at least one negative weight, meaning that they are extrapolations, rather than interpolations of the baseline shapes.  Generally, extrapolation is not possible with conventional design techniques. The Design-by-Morphing (DbM) response surface is highly sensitive to any changes in the values of the weights, which makes the DbM design space highly non-convex. We found that with each new epoch, small  changes in the weights  led to significant changes in the overall shape of the morphed draft tube/hub. This characteristic means that the DbM space needs to optimized with high precision where very small changes in the weights need to be considered. In particular, if we were to consider the value of each weight to be a discrete, rather than a continuous variable, then the design space would need to have a very large number of discrete variables. Since we are treating the weights here as continuous variables, the sensitive dependence of the morphed shape on the weight values means that most conventional optimization schemes, the search may get ``stuck'' at a local maximum and will fail to find the global maximum. Bayesian Optimization algorithms, including MixMOBO, that are designed to search for the {\it global} maximum of $C_{prm}$, will tend to  over-explore the region of the design space around a {\it } local maximum. This effect, however, is less pronounced for BO based schemes since they do not depend upon the gradient of the surrogate surfaces. MixMOBO also uses ``mutation'' to get out of the local maxima. Our numerical experiments with MixMOBO show that it tends to not get stuck at local maxima, and it finds the global maximum \cite{mixmobo}.

\section{Conclusion}
\label{sec:concl}

We have introduced a novel systematic shape optimization framework using the Design-by-Morphing (DbM) technique with a Mixed-variable Multi-Objective Bayesian Optimization (MixMOBO) algorithm. As a proof-of-concept of this optimization framework, we found the shape of a draft tube/hub for a hydrokinetic turbine such that the coefficient of pressure recovery $C_{prm}$ was maximized. Our Design-by-Morphing (DbM) technique creates novel shapes from interpolations and extrapolations of  pre-existing designs and/or new ones, and the shapes we create are free from designer's biases. Most existing design strategies cannot extrapolate among designs, and this limits their abilities to create radically different designs. 

The design space created with the Design-by-Morphing framework was searched by our novel MixBOBO Bayesian Optimization. This search method is especially useful when the property that is being optimized, in this case the $C_{prm}$, is costly to compute (or find experimentally). In these cases, it is necessary for the search algorithm to find the optimum design with as few evaluations of $C_{prm}$ as possible. Using MixMOBO, we successfully found a design with large $C_{prm}$ (certainly a local maximum in the design space, and possibley the global maximum)  with 30 epochs, or 200 evaluations of a design's $C_{prm}$.


The draft tube/hub design found here has a significantly better coefficient of mean pressure recovery than the  sharp-heel draft tube, which is the most commonly used draft tube for hydrokinetic turbines. More significantly, we have shown that the design framework used here, which can  easily be generalized to a number of engineering design optimization problems, with the current study providing a proof-of-concept for our optimization framework. DbM provides a methodology to create a bias and constraint free design space which also allows extrapolation from the existing designs and can yield radical shapes. MixMOBO provides a global optimization strategy to optimize optimize design spaces where evaluating each candidate design is extremely expensive, the design space contains mixed variables and/or multiple objectives, rendering conventional optimization techniques intractable. The framework can be used for optimizing expensive black-box shape optimization problems such as vertical-axis wind turbines \cite{2019APS..DFDQ14007S}, high-performance airfoils \cite{airfoils} and architected meta-materials \cite{mixmobo}.

\newpage

\newpage
\bibliographystyle{elsarticle-num-names} 
\bibliography{cas-refs}

\begin{thebibliography}{75}
\expandafter\ifx\csname natexlab\endcsname\relax\def\natexlab#1{#1}\fi
\providecommand{\url}[1]{\texttt{#1}}
\providecommand{\href}[2]{#2}
\providecommand{\path}[1]{#1}
\providecommand{\DOIprefix}{doi:}
\providecommand{\ArXivprefix}{arXiv:}
\providecommand{\URLprefix}{URL: }
\providecommand{\Pubmedprefix}{pmid:}
\providecommand{\doi}[1]{\href{http://dx.doi.org/#1}{\path{#1}}}
\providecommand{\Pubmed}[1]{\href{pmid:#1}{\path{#1}}}
\providecommand{\bibinfo}[2]{#2}
\ifx\xfnm\relax \def\xfnm[#1]{\unskip,\space#1}\fi
\bibitem[{eia(2022)}]{eiaenergyreview2021}
\bibinfo{title}{U.s. energy information administration monthly energy review,
  december 2021},
  \bibinfo{howpublished}{\url{https://www.eia.gov/totalenergy/data/monthly/archive/00352112.pdf}},
  \bibinfo{year}{2022}.
\bibitem[{Curtis et~al.(2017)Curtis, Levine, and Johnson}]{smallhydro_models}
\bibinfo{author}{T.~Curtis}, \bibinfo{author}{A.~Levine},
  \bibinfo{author}{K.~Johnson},
\newblock \bibinfo{title}{State models to incentivize and streamline small
  hydropower development},
\newblock \bibinfo{publisher}{National Renewable Energy Laboratory (NREL), U.S.
  Department of Energy}, \bibinfo{year}{2017}.
\bibitem[{of~Energy(2001)}]{smallhydro_demand}
\bibinfo{author}{U.~S.~D. of~Energy}, \bibinfo{title}{Hydropower vision},
  \bibinfo{howpublished}{\url{https://www.energy.gov/sites/default/files/2018/02/f49/Hydropower-Vision-021518.pdf}},
  \bibinfo{year}{2001}. \bibinfo{note}{Accessed 03-29-2022}.
\bibitem[{Bakken et~al.(2012)Bakken, Sundt, Ruud, and Harby}]{BAKKEN2012185}
\bibinfo{author}{T.~H. Bakken}, \bibinfo{author}{H.~Sundt},
  \bibinfo{author}{A.~Ruud}, \bibinfo{author}{A.~Harby},
\newblock \bibinfo{title}{Development of small versus large hydropower in
  norway– comparison of environmental impacts},
\newblock \bibinfo{journal}{Energy Procedia} \bibinfo{volume}{20}
  (\bibinfo{year}{2012}) \bibinfo{pages}{185--199}. \URLprefix
  \url{https://www.sciencedirect.com/science/article/pii/S1876610212007497}.
  \DOIprefix\doi{https://doi.org/10.1016/j.egypro.2012.03.019},
  \bibinfo{note}{technoport 2012 - Sharing Possibilities and 2nd Renewable
  Energy Research Conference (RERC2012)}.
\bibitem[{Couto~BA and Olden(2018)}]{smallhydro_ecology}
\bibinfo{author}{T.~Couto~BA}, \bibinfo{author}{J.~D. Olden},
\newblock \bibinfo{title}{Global proliferation of small hydropower plants -
  science and policy},
\newblock in: \bibinfo{booktitle}{Front Ecol Environ},
  volume~\bibinfo{volume}{16}, \bibinfo{year}{2018}, pp.
  \bibinfo{pages}{91--100}. \DOIprefix\doi{https://doi.org/10.1002/fee.1746}.
\bibitem[{Hall~G. et~al.(2006)Hall~G., S.~Reeves, Brizzee, Lee~D., Carroll~R.,
  and Sommers~L.}]{smallhydro_idaho}
\bibinfo{author}{D.~Hall~G.}, \bibinfo{author}{K.~S.~Reeves},
  \bibinfo{author}{J.~Brizzee}, \bibinfo{author}{R.~Lee~D.},
  \bibinfo{author}{G.~Carroll~R.}, \bibinfo{author}{G.~Sommers~L.},
\newblock \bibinfo{title}{Feasibility assessment of the water energy resources
  of the united states for new low power and small hydro classes of
  hydroelectric plants},
\newblock \bibinfo{publisher}{Idaho National Laboratory}, \bibinfo{year}{2006}.
\bibitem[{Kaltschmitt et~al.(2007)Kaltschmitt, Streicher, and Wiese}]{reneng}
\bibinfo{author}{M.~Kaltschmitt}, \bibinfo{author}{W.~Streicher},
  \bibinfo{author}{A.~Wiese}, \bibinfo{title}{Renewable Energy: Technology,
  Environment and Economics}, \bibinfo{publisher}{Springer, Berlin,
  Heidelberg}, \bibinfo{year}{2007}.
\bibitem[{Mulley(2004)}]{bernoulli}
\bibinfo{author}{R.~Mulley}, \bibinfo{title}{Flow of Industrial Fluids - Theory
  and Equations}, \bibinfo{publisher}{Taylor \& Francis Group, LLC},
  \bibinfo{year}{2004}.
\bibitem[{Warnick et~al.(1984)Warnick, Mayo, Carson, and Sheldon}]{sheldon}
\bibinfo{author}{C.~C. Warnick}, \bibinfo{author}{H.~A. Mayo},
  \bibinfo{author}{J.~L. Carson}, \bibinfo{author}{L.~H. Sheldon},
  \bibinfo{title}{Hydropower Engineering}, \bibinfo{publisher}{Prentice-Hall,
  Inc.}, \bibinfo{year}{1984}.
\bibitem[{Sheldon(1998)}]{sheldonopt}
\bibinfo{author}{L.~H. Sheldon},
\newblock \bibinfo{title}{Reviewing the approaches to hydro optimization},
\newblock \bibinfo{journal}{Hydro Review} \bibinfo{volume}{17}
  (\bibinfo{year}{1998}) \bibinfo{pages}{60--67}.
\bibitem[{Cervantes(2003)}]{cervantesthesis}
\bibinfo{author}{M.~J. Cervantes},
\newblock \bibinfo{title}{Effects of boundary conditions and unsteadiness on
  draft tube flow, ph.d. thesis},
\newblock \bibinfo{journal}{Division of Fluid Mechanics, Luleå University of
  Technology}  (\bibinfo{year}{2003}).
\bibitem[{Krivchenko(1994)}]{krivchenko}
\bibinfo{author}{G.~Krivchenko}, \bibinfo{title}{Hydraulic Machines: Turbines
  and Pumps}, \bibinfo{publisher}{CRC Press, Inc.}, \bibinfo{year}{1994}.
\bibitem[{Andersson(2009)}]{anderssonthesis}
\bibinfo{author}{U.~Andersson},
\newblock \bibinfo{title}{An experimental study of the flow in a sharp-heel
  kaplan draft tube, ph.d. thesis},
\newblock \bibinfo{journal}{Division of Fluid Mechanics, Luleå University of
  Technology}  (\bibinfo{year}{2009}).
\bibitem[{Nilsson(2006)}]{nilsson1}
\bibinfo{author}{H.~Nilsson},
\newblock \bibinfo{title}{3d numerical analysis of the unsteady turbulent
  swirling flow in a conical diffuser using fluent and openfoam},
\newblock in: \bibinfo{booktitle}{IAHR 2006}, \bibinfo{year}{2006}.
\bibitem[{Marjavaara(2006)}]{marjavaarathesis}
\bibinfo{author}{B.~D. Marjavaara},
\newblock \bibinfo{title}{Cfd driven optimization of hydraulic turbine draft
  tubes using surrogate models},
\newblock \bibinfo{journal}{Division of Fluid Mechanics, Luleå University of
  Technology}  (\bibinfo{year}{2006}).
\bibitem[{Dahlhaug(1997)}]{dahlhaug}
\bibinfo{author}{O.~G. Dahlhaug},
\newblock \bibinfo{title}{A study of swirl flow in draft tubes},
\newblock \bibinfo{publisher}{Norwegian University of Science and Technology
  Trondheim}, \bibinfo{year}{1997}.
\bibitem[{Gubin(1970)}]{gubin}
\bibinfo{author}{M.~Gubin}, \bibinfo{title}{Draft Tubes of Hydro-electric
  Stations}, \bibinfo{publisher}{Energiya Press, Moscow}, \bibinfo{year}{1970}.
\bibitem[{hol(2022)}]{holleforsen_2022}
\bibinfo{title}{Hölleforsen - vattenfall ab},
  \bibinfo{howpublished}{\url{https://powerplants.vattenfall.com/holleforsen/}},
  \bibinfo{year}{2022}. \bibinfo{note}{Accessed: 2022-03-28}.
\bibitem[{Dahlbäck(1996)}]{dahlback}
\bibinfo{author}{N.~Dahlbäck},
\newblock \bibinfo{title}{Redesign of sharp heel draft tube - results from
  tests in model and prototype},
\newblock in: \bibinfo{editor}{E.~Cabrera}, \bibinfo{editor}{V.~Espert},
  \bibinfo{editor}{F.~Martínez} (Eds.), \bibinfo{booktitle}{Proceedings of the
  XVIII IAHR Symposium on Hydraulic Macehinery and Cavitation},
  \bibinfo{publisher}{Springer-Science+Business Media, B.V.},
  \bibinfo{year}{1996}, pp. \bibinfo{pages}{985--993}.
\bibitem[{Marjavaara and Lundström(2003)}]{marjavaaralundstrom}
\bibinfo{author}{B.~Marjavaara}, \bibinfo{author}{T.~Lundström},
\newblock \bibinfo{title}{Automatic shape optimisation of a hydropower draft
  tube},
\newblock in: \bibinfo{editor}{E.~Cabrera}, \bibinfo{editor}{V.~Espert},
  \bibinfo{editor}{F.~Martínez} (Eds.), \bibinfo{booktitle}{ASME/JSME 4th
  Joint Fluids Summer Engineering Conference}, volume~\bibinfo{volume}{2},
  \bibinfo{year}{2003}, p. \bibinfo{pages}{1819–1824}.
\bibitem[{Daniels et~al.(2021)Daniels, Rahat, Tabor, Fieldsend, and
  Everson}]{daniels_1}
\bibinfo{author}{S.~J. Daniels}, \bibinfo{author}{A.~A.~M. Rahat},
  \bibinfo{author}{G.~R. Tabor}, \bibinfo{author}{J.~E. Fieldsend},
  \bibinfo{author}{R.~M. Everson},
\newblock \bibinfo{title}{Application of multi‐objective bayesian shape
  optimisation to a sharp‐heeled kaplan draft tube},
\newblock \bibinfo{journal}{Optimization and Engineering} \bibinfo{volume}{22}
  (\bibinfo{year}{2021}).
\bibitem[{Daniels et~al.(2018)Daniels, Rahat, Everson, Tabor, and
  Fieldsend}]{daniels_2}
\bibinfo{author}{S.~J. Daniels}, \bibinfo{author}{A.~A.~M. Rahat},
  \bibinfo{author}{R.~M. Everson}, \bibinfo{author}{G.~R. Tabor},
  \bibinfo{author}{J.~E. Fieldsend},
\newblock \bibinfo{title}{A suite of computationally expensive shape
  optimisation problems using computational fluid dynamics},
\newblock in: \bibinfo{editor}{A.~Auger}, \bibinfo{editor}{C.~M. Fonseca},
  \bibinfo{editor}{N.~Louren{\c{c}}o}, \bibinfo{editor}{P.~Machado},
  \bibinfo{editor}{L.~Paquete}, \bibinfo{editor}{D.~Whitley} (Eds.),
  \bibinfo{booktitle}{Parallel Problem Solving from Nature -- PPSN XV},
  \bibinfo{publisher}{Springer International Publishing},
  \bibinfo{address}{Cham}, \bibinfo{year}{2018}, pp. \bibinfo{pages}{296--307}.
\bibitem[{Eisinger and Ruprecht(2002)}]{eisinger}
\bibinfo{author}{R.~Eisinger}, \bibinfo{author}{A.~Ruprecht},
\newblock \bibinfo{title}{Automatic shape optimisation of hydro turbine
  components based on cfd},
\newblock \bibinfo{journal}{TASK quarterly: Scietific Bulletin of Academic
  Computing Centre Gdańsk} \bibinfo{volume}{6} (\bibinfo{year}{2002})
  \bibinfo{pages}{101–111}.
\bibitem[{McNabb et~al.(2014)McNabb, Devals, Kyriacou, Murry, and
  Mullins}]{mcnabb}
\bibinfo{author}{J.~McNabb}, \bibinfo{author}{C.~Devals},
  \bibinfo{author}{S.~Kyriacou}, \bibinfo{author}{N.~Murry},
  \bibinfo{author}{B.~Mullins},
\newblock \bibinfo{title}{Cfd based draft-tube hydraulic design optimisation},
\newblock in: \bibinfo{booktitle}{27th IAHR symposium on hydraulic machinery
  and systems}, \bibinfo{year}{2014}.
\bibitem[{Sobester and Barrett(2008)}]{sobester_barrett_2008}
\bibinfo{author}{A.~Sobester}, \bibinfo{author}{T.~Barrett},
\newblock \bibinfo{title}{Quest for a truly parsimonious airfoil
  parameterization scheme},
\newblock \bibinfo{journal}{The 26th Congress of ICAS and 8th AIAA ATIO}
  (\bibinfo{year}{2008}). \DOIprefix\doi{10.2514/6.2008-8879}.
\bibitem[{Sripawadkul et~al.(2010)Sripawadkul, Padulo, and
  Guenov}]{sripawadkul_comparison_2010}
\bibinfo{author}{V.~Sripawadkul}, \bibinfo{author}{M.~Padulo},
  \bibinfo{author}{M.~Guenov},
\newblock \bibinfo{title}{A {Comparison} of {Airfoil} {Shape}
  {Parameterization} {Techniques} for {Early} {Design} {Optimization}},
\newblock in: \bibinfo{booktitle}{13th {AIAA}/{ISSMO} {Multidisciplinary}
  {Analysis} {Optimization} {Conference}}, \bibinfo{publisher}{American
  Institute of Aeronautics and Astronautics}, \bibinfo{address}{Fort Worth,
  Texas}, \bibinfo{year}{2010}. \URLprefix
  \url{https://arc.aiaa.org/doi/10.2514/6.2010-9050}.
  \DOIprefix\doi{10.2514/6.2010-9050}.
\bibitem[{Masters et~al.(2015)Masters, Taylor, Rendall, Allen, and
  Poole}]{masters_review_2015}
\bibinfo{author}{D.~A. Masters}, \bibinfo{author}{N.~J. Taylor},
  \bibinfo{author}{T.~Rendall}, \bibinfo{author}{C.~B. Allen},
  \bibinfo{author}{D.~J. Poole},
\newblock \bibinfo{title}{Review of {Aerofoil} {Parameterisation} {Methods} for
  {Aerodynamic} {Shape} {Optimisation}},
\newblock in: \bibinfo{booktitle}{53rd {AIAA} {Aerospace} {Sciences}
  {Meeting}}, \bibinfo{publisher}{American Institute of Aeronautics and
  Astronautics}, \bibinfo{address}{Kissimmee, Florida}, \bibinfo{year}{2015}.
  \URLprefix \url{https://arc.aiaa.org/doi/10.2514/6.2015-0761}.
  \DOIprefix\doi{10.2514/6.2015-0761}.
\bibitem[{Moazam~Sheikh et~al.(2017)Moazam~Sheikh, Shabbir, Ahmed, Waseem, and
  Sheikh}]{moazam2017computational}
\bibinfo{author}{H.~Moazam~Sheikh}, \bibinfo{author}{Z.~Shabbir},
  \bibinfo{author}{H.~Ahmed}, \bibinfo{author}{M.~H. Waseem},
  \bibinfo{author}{M.~Z. Sheikh},
\newblock \bibinfo{title}{Computational fluid dynamics analysis of a modified
  savonius rotor and optimization using response surface methodology},
\newblock \bibinfo{journal}{Wind Engineering} \bibinfo{volume}{41}
  (\bibinfo{year}{2017}) \bibinfo{pages}{285--296}.
\bibitem[{Jameson(1988)}]{discrete_method}
\bibinfo{author}{A.~Jameson},
\newblock \bibinfo{title}{Aerodynamic design via control theory},
\newblock \bibinfo{journal}{Journal of Scientific Computing}
  \bibinfo{volume}{3} (\bibinfo{year}{1988}) \bibinfo{pages}{233–260}.
  \URLprefix \url{https://link.springer.com/article/10.1007/BF01061285}.
  \DOIprefix\doi{10.1007/bf01061285}.
\bibitem[{Samareh(2001)}]{survey_parameterization}
\bibinfo{author}{J.~A. Samareh},
\newblock \bibinfo{title}{Survey of shape parameterization techniques for
  high-fidelity multidisciplinary shape optimization},
\newblock \bibinfo{journal}{AIAA Journal} \bibinfo{volume}{39}
  (\bibinfo{year}{2001}) \bibinfo{pages}{877–884}. \URLprefix
  \url{https://ui.adsabs.harvard.edu/abs/2001AIAAJ..39..877S/abstract}.
  \DOIprefix\doi{10.2514/2.1391}.
\bibitem[{Sanaye and Hassanzadeh(2014)}]{sanaye_multi-objective_2014}
\bibinfo{author}{S.~Sanaye}, \bibinfo{author}{A.~Hassanzadeh},
\newblock \bibinfo{title}{Multi-objective optimization of airfoil shape for
  efficiency improvement and noise reduction in small wind turbines},
\newblock \bibinfo{journal}{Journal of Renewable and Sustainable Energy}
  \bibinfo{volume}{6} (\bibinfo{year}{2014}) \bibinfo{pages}{053105}.
  \URLprefix \url{http://aip.scitation.org/doi/10.1063/1.4895528}.
  \DOIprefix\doi{10.1063/1.4895528}.
\bibitem[{Han and Zingg(2014)}]{han_adaptive_2014}
\bibinfo{author}{X.~Han}, \bibinfo{author}{D.~W. Zingg},
\newblock \bibinfo{title}{An adaptive geometry parametrization for aerodynamic
  shape optimization},
\newblock \bibinfo{journal}{Optimization and Engineering} \bibinfo{volume}{15}
  (\bibinfo{year}{2014}) \bibinfo{pages}{69--91}. \URLprefix
  \url{http://link.springer.com/10.1007/s11081-013-9213-y}.
  \DOIprefix\doi{10.1007/s11081-013-9213-y}.
\bibitem[{Schramm et~al.(1995)Schramm, Pilkey, DeVries, and Zebrowski}]{nurbs}
\bibinfo{author}{U.~Schramm}, \bibinfo{author}{W.~D. Pilkey},
  \bibinfo{author}{R.~I. DeVries}, \bibinfo{author}{M.~P. Zebrowski},
\newblock \bibinfo{title}{Shape design for thin-walled beam cross sections
  using rational b splines},
\newblock \bibinfo{journal}{AIAA Journal} \bibinfo{volume}{33}
  (\bibinfo{year}{1995}) \bibinfo{pages}{2205–2211}.
  \DOIprefix\doi{10.2514/3.12870}.
\bibitem[{Sederberg and Parry(1986)}]{FFD1}
\bibinfo{author}{T.~W. Sederberg}, \bibinfo{author}{S.~R. Parry},
\newblock \bibinfo{title}{Free-form deformation of solid geometric models},
\newblock \bibinfo{journal}{ACM SIGGRAPH Computer Graphics}
  \bibinfo{volume}{20} (\bibinfo{year}{1986}) \bibinfo{pages}{151–160}.
  \URLprefix \url{https://dl.acm.org/doi/epdf/10.1145/15886.15903}.
  \DOIprefix\doi{10.1145/15886.15903}.
\bibitem[{Lamousin and Waggenspack(1994)}]{FFD2}
\bibinfo{author}{H.~Lamousin}, \bibinfo{author}{N.~Waggenspack},
\newblock \bibinfo{title}{Nurbs-based free-form deformations},
\newblock \bibinfo{journal}{IEEE Computer Graphics and Applications}
  \bibinfo{volume}{14} (\bibinfo{year}{1994}) \bibinfo{pages}{59–65}.
  \URLprefix \url{https://ieeexplore.ieee.org/document/329096}.
  \DOIprefix\doi{10.1109/38.329096}.
\bibitem[{J.~Toal et~al.(2010)J.~Toal, Bressloff, Keane, and E.~Holden}]{POD1}
\bibinfo{author}{D.~J. J.~Toal}, \bibinfo{author}{N.~W. Bressloff},
  \bibinfo{author}{A.~J. Keane}, \bibinfo{author}{C.~M. E.~Holden},
\newblock \bibinfo{title}{Geometric filtration using proper orthogonal
  decomposition for aerodynamic design optimization},
\newblock \bibinfo{journal}{AIAA Journal} \bibinfo{volume}{48}
  (\bibinfo{year}{2010}) \bibinfo{pages}{916–928}.
  \DOIprefix\doi{10.2514/1.41420}.
\bibitem[{Ghoman et~al.(2012)Ghoman, Wang, Chen, and Kapania}]{POD2}
\bibinfo{author}{S.~Ghoman}, \bibinfo{author}{Z.~Wang},
  \bibinfo{author}{P.~Chen}, \bibinfo{author}{R.~Kapania},
\newblock \bibinfo{title}{A pod-based reduced order design scheme for shape
  optimization of air vehicles},
\newblock \bibinfo{journal}{53rd AIAA/ASME/ASCE/AHS/ASC Structures, Structural
  Dynamics and Materials Conference<BR>20th AIAA/ASME/AHS Adaptive Structures
  Conference<BR>14th AIAA}  (\bibinfo{year}{2012}).
  \DOIprefix\doi{10.2514/6.2012-1808}.
\bibitem[{Hicks and Henne(1978)}]{hicks_wing_1978}
\bibinfo{author}{R.~M. Hicks}, \bibinfo{author}{P.~A. Henne},
\newblock \bibinfo{title}{Wing {Design} by {Numerical} {Optimization}},
\newblock \bibinfo{journal}{Journal of Aircraft} \bibinfo{volume}{15}
  (\bibinfo{year}{1978}) \bibinfo{pages}{407--412}. \URLprefix
  \url{https://arc.aiaa.org/doi/10.2514/3.58379}.
  \DOIprefix\doi{10.2514/3.58379}.
\bibitem[{Kulfan and Bussoletti(2006)}]{kulfan_fundamental_2006}
\bibinfo{author}{B.~Kulfan}, \bibinfo{author}{J.~Bussoletti},
\newblock \bibinfo{title}{"{Fundamental}" {Parameteric} {Geometry}
  {Representations} for {Aircraft} {Component} {Shapes}},
\newblock in: \bibinfo{booktitle}{11th {AIAA}/{ISSMO} {Multidisciplinary}
  {Analysis} and {Optimization} {Conference}}, \bibinfo{publisher}{American
  Institute of Aeronautics and Astronautics}, \bibinfo{address}{Portsmouth,
  Virginia}, \bibinfo{year}{2006}. \URLprefix
  \url{http://arc.aiaa.org/doi/10.2514/6.2006-6948}.
  \DOIprefix\doi{10.2514/6.2006-6948}.
\bibitem[{Akram and Kim(2021)}]{akram_cfd_2021}
\bibinfo{author}{M.~T. Akram}, \bibinfo{author}{M.-H. Kim},
\newblock \bibinfo{title}{{CFD} {Analysis} and {Shape} {Optimization} of
  {Airfoils} {Using} {Class} {Shape} {Transformation} and {Genetic}
  {Algorithm}—{Part} {I}},
\newblock \bibinfo{journal}{Applied Sciences} \bibinfo{volume}{11}
  (\bibinfo{year}{2021}) \bibinfo{pages}{3791}. \URLprefix
  \url{https://www.mdpi.com/2076-3417/11/9/3791}.
  \DOIprefix\doi{10.3390/app11093791}.
\bibitem[{Sheikh et~al.(2022)Sheikh, Lee, Wang, and Marcus}]{airfoils}
\bibinfo{author}{H.~M. Sheikh}, \bibinfo{author}{S.~Lee},
  \bibinfo{author}{J.~Wang}, \bibinfo{author}{P.~S. Marcus},
  \bibinfo{title}{Airfoil optimization using design-by-morphing},
  \bibinfo{year}{2022}. \URLprefix \url{https://arxiv.org/abs/2207.11448}.
  \DOIprefix\doi{10.48550/ARXIV.2207.11448}.
\bibitem[{{Oh} et~al.(2018){Oh}, {Jiang}, {Jiang}, and
  {Marcus}}]{2018CompM6223O}
\bibinfo{author}{S.~{Oh}}, \bibinfo{author}{C.-H. {Jiang}},
  \bibinfo{author}{C.~{Jiang}}, \bibinfo{author}{P.~S. {Marcus}},
\newblock \bibinfo{title}{{Finding the optimal shape of the
  leading-and-trailing car of a high-speed train using design-by-morphing}},
\newblock \bibinfo{journal}{Computational Mechanics} \bibinfo{volume}{62}
  (\bibinfo{year}{2018}) \bibinfo{pages}{23--45}.
  \DOIprefix\doi{10.1007/s00466-017-1482-4}.
\bibitem[{{Sheikh} and {Marcus}(2019)}]{2019APS..DFDQ14007S}
\bibinfo{author}{H.~M. {Sheikh}}, \bibinfo{author}{P.~S. {Marcus}},
\newblock \bibinfo{title}{{Vertical Axis Wind Turbine Design Using
  Design-by-Morphing and Bayesian Optimization}},
\newblock in: \bibinfo{booktitle}{APS Division of Fluid Dynamics Meeting
  Abstracts}, APS Meeting Abstracts, \bibinfo{year}{2019}, p.
  \bibinfo{pages}{Q14.007}.
\bibitem[{Schramm et~al.(2018)Schramm, Stoevesandt, and
  Peinke}]{computation6010005}
\bibinfo{author}{M.~Schramm}, \bibinfo{author}{B.~Stoevesandt},
  \bibinfo{author}{J.~Peinke},
\newblock \bibinfo{title}{Optimization of airfoils using the adjoint approach
  and the influence of adjoint turbulent viscosity},
\newblock \bibinfo{journal}{Computation} \bibinfo{volume}{6}
  (\bibinfo{year}{2018}). \URLprefix
  \url{https://www.mdpi.com/2079-3197/6/1/5}.
  \DOIprefix\doi{10.3390/computation6010005}.
\bibitem[{{Chen} et~al.(2007){Chen}, {Shapiro}, {Suresh}, and
  {Tsukanov}}]{2007IJNME..71..313C}
\bibinfo{author}{J.~{Chen}}, \bibinfo{author}{V.~{Shapiro}},
  \bibinfo{author}{K.~{Suresh}}, \bibinfo{author}{I.~{Tsukanov}},
\newblock \bibinfo{title}{{Shape optimization with topological changes and
  parametric control}},
\newblock \bibinfo{journal}{International Journal for Numerical Methods in
  Engineering} \bibinfo{volume}{71} (\bibinfo{year}{2007})
  \bibinfo{pages}{313--346}. \DOIprefix\doi{10.1002/nme.1943}.
\bibitem[{{Zhang} et~al.(1995){Zhang}, {Beckers}, and
  {Fleury}}]{1995IJNME..38.2283Z}
\bibinfo{author}{W.-H. {Zhang}}, \bibinfo{author}{P.~{Beckers}},
  \bibinfo{author}{C.~{Fleury}},
\newblock \bibinfo{title}{{A unified parametric design approach to structural
  shape optimization}},
\newblock \bibinfo{journal}{International Journal for Numerical Methods in
  Engineering} \bibinfo{volume}{38} (\bibinfo{year}{1995})
  \bibinfo{pages}{2283--2292}. \DOIprefix\doi{10.1002/nme.1620381309}.
\bibitem[{{Hughes} et~al.(2005){Hughes}, {Cottrell}, and
  {Bazilevs}}]{2005CMAME.194.4135H}
\bibinfo{author}{T.~J.~R. {Hughes}}, \bibinfo{author}{J.~A. {Cottrell}},
  \bibinfo{author}{Y.~{Bazilevs}},
\newblock \bibinfo{title}{{Isogeometric analysis: CAD, finite elements, NURBS,
  exact geometry and mesh refinement}},
\newblock \bibinfo{journal}{Computer Methods in Applied Mechanics and
  Engineering} \bibinfo{volume}{194} (\bibinfo{year}{2005})
  \bibinfo{pages}{4135--4195}. \DOIprefix\doi{10.1016/j.cma.2004.10.008}.
\bibitem[{{Shyy} et~al.(2001){Shyy}, {Papila}, {Vaidyanathan}, and
  {Tucker}}]{2001PrAeS..37...59S}
\bibinfo{author}{W.~{Shyy}}, \bibinfo{author}{N.~{Papila}},
  \bibinfo{author}{R.~{Vaidyanathan}}, \bibinfo{author}{K.~{Tucker}},
\newblock \bibinfo{title}{{Global design optimization for aerodynamics and
  rocket propulsion components}},
\newblock \bibinfo{journal}{Progress in Aerospace Sciences}
  \bibinfo{volume}{37} (\bibinfo{year}{2001}) \bibinfo{pages}{59--118}.
  \DOIprefix\doi{10.1016/S0376-0421(01)00002-1}.
\bibitem[{{Wang} et~al.(2011){Wang}, {Hirsch}, {Kang}, and
  {Lacor}}]{2011CMAME.200..883W}
\bibinfo{author}{X.~D. {Wang}}, \bibinfo{author}{C.~{Hirsch}},
  \bibinfo{author}{S.~{Kang}}, \bibinfo{author}{C.~{Lacor}},
\newblock \bibinfo{title}{{Multi-objective optimization of turbomachinery using
  improved NSGA-II and approximation model}},
\newblock \bibinfo{journal}{Computer Methods in Applied Mechanics and
  Engineering} \bibinfo{volume}{200} (\bibinfo{year}{2011})
  \bibinfo{pages}{883--895}. \DOIprefix\doi{10.1016/j.cma.2010.11.014}.
\bibitem[{Fang and Li(2015)}]{FANG1329}
\bibinfo{author}{L.~Fang}, \bibinfo{author}{X.~Li},
\newblock \bibinfo{title}{Design optimization of unsteady airfoils with
  continuous adjoint method},
\newblock \bibinfo{journal}{Applied Mathematics and Mechanics}
  \bibinfo{volume}{36} (\bibinfo{year}{2015}) \bibinfo{pages}{1329}. \URLprefix
  \url{https://www.amm.shu.edu.cn/CN/abstract/article_16192.shtml}.
  \DOIprefix\doi{10.1007/s10483-015-2010-9}.
\bibitem[{Sheikh and Marcus(2022)}]{mixmobo}
\bibinfo{author}{H.~M. Sheikh}, \bibinfo{author}{P.~S. Marcus},
  \bibinfo{title}{Bayesian optimization for multi-objective mixed-variable
  problems}, \bibinfo{year}{2022}. \href{http://arxiv.org/abs/2201.12767}{{\tt
  arXiv:2201.12767}}.
\bibitem[{Brochu et~al.(2010)Brochu, Cora, and de~Freitas}]{brochu2010tutorial}
\bibinfo{author}{E.~Brochu}, \bibinfo{author}{V.~M. Cora},
  \bibinfo{author}{N.~de~Freitas}, \bibinfo{title}{A tutorial on bayesian
  optimization of expensive cost functions, with application to active user
  modeling and hierarchical reinforcement learning}, \bibinfo{year}{2010}.
  \href{http://arxiv.org/abs/1012.2599}{{\tt arXiv:1012.2599}}.
\bibitem[{Williams and Rasmussen(2006)}]{williams2006gaussian}
\bibinfo{author}{C.~K.~I. Williams}, \bibinfo{author}{C.~E. Rasmussen},
  \bibinfo{title}{Gaussian processes for machine learning},
  volume~\bibinfo{volume}{2}, \bibinfo{publisher}{MIT press Cambridge, MA},
  \bibinfo{year}{2006}.
\bibitem[{Frazier and Wang(2015)}]{frazier2015}
\bibinfo{author}{P.~I. Frazier}, \bibinfo{author}{J.~Wang},
\newblock \bibinfo{title}{Bayesian optimization for materials design},
\newblock \bibinfo{journal}{Springer Series in Materials Science}
  (\bibinfo{year}{2015}) \bibinfo{pages}{45–75}.
  \DOIprefix\doi{10.1007/978-3-319-23871-5_3}.
\bibitem[{Chen et~al.(2018)Chen, Skouras, Zhu, and
  Matusik}]{chen2018computational}
\bibinfo{author}{D.~Chen}, \bibinfo{author}{M.~Skouras},
  \bibinfo{author}{B.~Zhu}, \bibinfo{author}{W.~Matusik},
\newblock \bibinfo{title}{Computational discovery of extremal microstructure
  families},
\newblock \bibinfo{journal}{Science Advances} \bibinfo{volume}{4}
  (\bibinfo{year}{2018}) \bibinfo{pages}{eaao7005}.
\bibitem[{Chen et~al.(2019)Chen, Watts, Jackson, Smith, Tortorelli, and
  Spadaccini}]{chen2019stiff}
\bibinfo{author}{W.~Chen}, \bibinfo{author}{S.~Watts}, \bibinfo{author}{J.~A.
  Jackson}, \bibinfo{author}{W.~L. Smith}, \bibinfo{author}{D.~A. Tortorelli},
  \bibinfo{author}{C.~M. Spadaccini},
\newblock \bibinfo{title}{Stiff isotropic lattices beyond the maxwell
  criterion},
\newblock \bibinfo{journal}{Science Advances} \bibinfo{volume}{5}
  (\bibinfo{year}{2019}) \bibinfo{pages}{eaaw1937}.
\bibitem[{Shaw et~al.(2019)Shaw, Sun, Portela, Barranco, Greer, and
  Hopkins}]{shaw2019computationally}
\bibinfo{author}{L.~A. Shaw}, \bibinfo{author}{F.~Sun}, \bibinfo{author}{C.~M.
  Portela}, \bibinfo{author}{R.~I. Barranco}, \bibinfo{author}{J.~R. Greer},
  \bibinfo{author}{J.~B. Hopkins},
\newblock \bibinfo{title}{Computationally efficient design of directionally
  compliant metamaterials},
\newblock \bibinfo{journal}{Nature Communications} \bibinfo{volume}{10}
  (\bibinfo{year}{2019}) \bibinfo{pages}{1--13}.
\bibitem[{Song et~al.(2019)Song, Wang, Zhou, Fan, Yu, Lu, and
  Li}]{song2019topology}
\bibinfo{author}{J.~Song}, \bibinfo{author}{Y.~Wang},
  \bibinfo{author}{W.~Zhou}, \bibinfo{author}{R.~Fan}, \bibinfo{author}{B.~Yu},
  \bibinfo{author}{Y.~Lu}, \bibinfo{author}{L.~Li},
\newblock \bibinfo{title}{Topology optimization-guided lattice composites and
  their mechanical characterizations},
\newblock \bibinfo{journal}{Composites Part B: Engineering}
  \bibinfo{volume}{160} (\bibinfo{year}{2019}) \bibinfo{pages}{402--411}.
\bibitem[{Vangelatos et~al.(2021)Vangelatos, Sheikh, Marcus, Grigoropoulos,
  Lopez, Flamourakis, and Farsari}]{sadvours}
\bibinfo{author}{Z.~Vangelatos}, \bibinfo{author}{H.~M. Sheikh},
  \bibinfo{author}{P.~S. Marcus}, \bibinfo{author}{C.~P. Grigoropoulos},
  \bibinfo{author}{V.~Z. Lopez}, \bibinfo{author}{G.~Flamourakis},
  \bibinfo{author}{M.~Farsari},
\newblock \bibinfo{title}{Strength through defects: A novel bayesian approach
  for the optimization of architected materials},
\newblock \bibinfo{journal}{Science Advances} \bibinfo{volume}{7}
  (\bibinfo{year}{2021}). \DOIprefix\doi{10.1126/sciadv.abk2218}.
\bibitem[{Snoek et~al.(2012)Snoek, Larochelle, and Adams}]{snoek2012practical}
\bibinfo{author}{J.~Snoek}, \bibinfo{author}{H.~Larochelle},
  \bibinfo{author}{R.~P. Adams}, \bibinfo{title}{Practical bayesian
  optimization of machine learning algorithms}, \bibinfo{year}{2012}.
  \href{http://arxiv.org/abs/1206.2944}{{\tt arXiv:1206.2944}}.
\bibitem[{Chen et~al.(2018)Chen, Huang, Wang, Antonoglou, Schrittwieser,
  Silver, and de~Freitas}]{alphago2018}
\bibinfo{author}{Y.~Chen}, \bibinfo{author}{A.~Huang},
  \bibinfo{author}{Z.~Wang}, \bibinfo{author}{I.~Antonoglou},
  \bibinfo{author}{J.~Schrittwieser}, \bibinfo{author}{D.~Silver},
  \bibinfo{author}{N.~de~Freitas},
\newblock \bibinfo{title}{Bayesian optimization in alphago},
\newblock \bibinfo{journal}{CoRR} \bibinfo{volume}{abs/1812.06855}
  (\bibinfo{year}{2018}). \href{http://arxiv.org/abs/1812.06855}{{\tt
  arXiv:1812.06855}}.
\bibitem[{Oh et~al.(2018)Oh, Gavves, and Welling}]{pmlr-v80-oh18a}
\bibinfo{author}{C.~Oh}, \bibinfo{author}{E.~Gavves},
  \bibinfo{author}{M.~Welling},
\newblock \bibinfo{title}{{BOCK} : {B}ayesian optimization with cylindrical
  kernels},
\newblock in: \bibinfo{booktitle}{Proceedings of the 35th International
  Conference on Machine Learning}, volume~\bibinfo{volume}{80} of
  \textit{\bibinfo{series}{Proceedings of Machine Learning Research}},
  \bibinfo{publisher}{PMLR}, \bibinfo{year}{2018}, pp.
  \bibinfo{pages}{3868--3877}.
\bibitem[{Pyzer-Knapp(2018)}]{articledrug}
\bibinfo{author}{E.~Pyzer-Knapp},
\newblock \bibinfo{title}{Bayesian optimization for accelerated drug
  discovery},
\newblock \bibinfo{journal}{IBM Journal of Research and Development}
  \bibinfo{volume}{PP} (\bibinfo{year}{2018}) \bibinfo{pages}{1--1}.
  \DOIprefix\doi{10.1147/JRD.2018.2881731}.
\bibitem[{Korovina et~al.(2020)Korovina, Xu, Kandasamy, Neiswanger, Poczos,
  Schneider, and Xing}]{pmlr-v108-korovina20a}
\bibinfo{author}{K.~Korovina}, \bibinfo{author}{S.~Xu},
  \bibinfo{author}{K.~Kandasamy}, \bibinfo{author}{W.~Neiswanger},
  \bibinfo{author}{B.~Poczos}, \bibinfo{author}{J.~Schneider},
  \bibinfo{author}{E.~Xing},
\newblock \bibinfo{title}{Chembo: Bayesian optimization of small organic
  molecules with synthesizable recommendations},
\newblock in: \bibinfo{booktitle}{Proceedings of the Twenty Third International
  Conference on Artificial Intelligence and Statistics}, volume
  \bibinfo{volume}{108} of \textit{\bibinfo{series}{Proceedings of Machine
  Learning Research}}, \bibinfo{publisher}{PMLR}, \bibinfo{year}{2020}, pp.
  \bibinfo{pages}{3393--3403}.
\bibitem[{Krause et~al.(2008)Krause, Singh, and Guestrin}]{JMLR:v9:krause08a}
\bibinfo{author}{A.~Krause}, \bibinfo{author}{A.~Singh},
  \bibinfo{author}{C.~Guestrin},
\newblock \bibinfo{title}{Near-optimal sensor placements in gaussian processes:
  Theory, efficient algorithms and empirical studies},
\newblock \bibinfo{journal}{Journal of Machine Learning Research}
  \bibinfo{volume}{9} (\bibinfo{year}{2008}) \bibinfo{pages}{235--284}.
\bibitem[{Cervantes et~al.(2005)Cervantes, Engström, and
  Gustavsson}]{turbine99_iii}
\bibinfo{author}{M.~Cervantes}, \bibinfo{author}{T.~Engström},
  \bibinfo{author}{L.~Gustavsson},
\newblock in: \bibinfo{booktitle}{Proceedings of the third IAHR/ERCOFTAC
  Workshop on draft tube flows: Turbine-99 III}, \bibinfo{publisher}{Division
  of Fluid Mechanics, Luleå University of Technology}, \bibinfo{year}{2005}.
\bibitem[{Mulu(2012)}]{muluthesis}
\bibinfo{author}{B.~G. Mulu},
\newblock \bibinfo{title}{An experimental and numerical investigation of a
  kaplan turbine model, ph.d. thesis},
\newblock \bibinfo{journal}{Division of Fluid and Experimental Mechanics,
  Luleå University of Technology}  (\bibinfo{year}{2012}).
\bibitem[{Nilsson et~al.(2009)Nilsson, Muntean, and Susan-Resiga}]{nilsson2}
\bibinfo{author}{H.~Nilsson}, \bibinfo{author}{S.~Muntean},
  \bibinfo{author}{R.~F. Susan-Resiga},
\newblock \bibinfo{title}{Evaluation of openfoam for cfd of turbulent flow in
  water turbines},
\newblock in: \bibinfo{booktitle}{IAHR 2009}, \bibinfo{year}{2009}.
\bibitem[{Gebart et~al.(2000)Gebart, Gustavsson, and Karlsson}]{turbine99_i}
\bibinfo{author}{B.~R. Gebart}, \bibinfo{author}{L.~Gustavsson},
  \bibinfo{author}{R.~Karlsson},
\newblock in: \bibinfo{booktitle}{Turbine-99: Workshop on Draft Tube Flow},
  \bibinfo{publisher}{Division of Fluid Mechanics, Luleå University of
  Technology}, \bibinfo{year}{2000}.
\bibitem[{Engström et~al.(2001)Engström, Gustavsson, and
  Karlsson}]{turbine99_ii}
\bibinfo{author}{T.~Engström}, \bibinfo{author}{L.~Gustavsson},
  \bibinfo{author}{R.~Karlsson},
\newblock in: \bibinfo{booktitle}{Proceedings of Turbine-99 - Workshop 2: The
  second ERCOFTAC Workshop on Draft Tube Flow}, \bibinfo{publisher}{Division of
  Fluid Mechanics, Luleå University of Technology}, \bibinfo{year}{2001}.
\bibitem[{Mulu et~al.(2012)Mulu, Jonsson, and Cervantes}]{MuluJonsson2012}
\bibinfo{author}{B.~Mulu}, \bibinfo{author}{P.~Jonsson},
  \bibinfo{author}{M.~Cervantes},
\newblock \bibinfo{title}{Experimental investigation of a kaplan draft tube –
  part i: Best efficiency point},
\newblock \bibinfo{journal}{Division of Fluid Mechanics, Luleå University of
  Technology}  (\bibinfo{year}{2012}).
\bibitem[{Geuzaine and Remacle(2022)}]{gmsh}
\bibinfo{author}{C.~Geuzaine}, \bibinfo{author}{J.-F. Remacle},
  \bibinfo{title}{Gmsh}, \bibinfo{year}{2022}. \URLprefix
  \url{http://gmsh.info/}.
\bibitem[{Wu et~al.(2012)Wu, Liu, Dou, Wu, and Chen}]{wu}
\bibinfo{author}{Y.~Wu}, \bibinfo{author}{S.~Liu}, \bibinfo{author}{H.-S. Dou},
  \bibinfo{author}{S.~Wu}, \bibinfo{author}{T.~Chen},
\newblock \bibinfo{title}{Numerical prediction and similarity study of pressure
  fluctuation in a prototype kaplan turbine and the model turbine},
\newblock \bibinfo{journal}{Computers \& Fluids} \bibinfo{volume}{56}
  (\bibinfo{year}{2012}) \bibinfo{pages}{128–142}.
  \DOIprefix\doi{10.1016/j.compfluid.2011.12.005}.
\bibitem[{Brochu et~al.(2011)Brochu, Hoffman, and
  de~Freitas}]{brochu2011portfolio}
\bibinfo{author}{E.~Brochu}, \bibinfo{author}{M.~W. Hoffman},
  \bibinfo{author}{N.~de~Freitas}, \bibinfo{title}{Portfolio allocation for
  bayesian optimization}, \bibinfo{year}{2011}.
  \href{http://arxiv.org/abs/1009.5419}{{\tt arXiv:1009.5419}}.
\bibitem[{Tu\v{s}ar et~al.(2019)Tu\v{s}ar, Brockhoff, and Hansen}]{HAL}
\bibinfo{author}{T.~Tu\v{s}ar}, \bibinfo{author}{D.~Brockhoff},
  \bibinfo{author}{N.~Hansen},
\newblock \bibinfo{title}{Mixed-integer benchmark problems for single- and
  bi-objective optimization},
\newblock in: \bibinfo{booktitle}{Proceedings of the Genetic and Evolutionary
  Computation Conference}, GECCO '19, \bibinfo{publisher}{Association for
  Computing Machinery}, \bibinfo{address}{New York, NY, USA},
  \bibinfo{year}{2019}, p. \bibinfo{pages}{718–726}. \URLprefix
  \url{https://doi.org/10.1145/3321707.3321868}.
  \DOIprefix\doi{10.1145/3321707.3321868}.

\end{thebibliography}

\end{document}